\documentclass[journal]{IEEEtran}
\IEEEoverridecommandlockouts
\usepackage{cite}
\usepackage{amsmath,amssymb,amsfonts}
\usepackage{algorithmic}
\usepackage{graphicx}
 \usepackage{grffile} 
\usepackage{textcomp}
\usepackage{xcolor}
\usepackage{bm}
\usepackage{comment}
\usepackage{breqn}
\usepackage{subcaption}
\usepackage{booktabs}
\def\BibTeX{{\rm B\kern-.05em{\sc i\kern-.025em b}\kern-.08em
    T\kern-.1667em\lower.7ex\hbox{E}\kern-.125emX}}

\begin{document}

\title{Vital Signs Estimation Using a 26 GHz Multi-Beam Communication Testbed
}


\author{Miquel Sellés Valls,~\IEEEmembership{Student Member,~IEEE}, Sofie Pollin,~\IEEEmembership{Senior Member,~IEEE}, Ying Wang,~\IEEEmembership{Member,~IEEE}, Rizqi Hersyandika,~\IEEEmembership{Student Member,~IEEE}, Andre Kokkeler,~\IEEEmembership{Senior Member,~IEEE}, Yang Miao,~\IEEEmembership{Member,~IEEE}

\thanks{M.S. Valls, Y. Wang, A. Kokkeler, and Y. Miao are with the Department of Electrical Engineering, University of Twente, the Netherlands. S. Pollin, R. Hersyandika and Y. Miao are with the Department of Electrical Engineering, KU Leuven, Belgium. 
}
}



\maketitle

\begin{abstract}

This paper presents a novel pipeline for vital sign monitoring using a 26 GHz multi-beam communication testbed. In context of Joint Communication and Sensing (JCAS), the advanced communication capability at millimeter-wave bands is comparable to the radio resource of radars and is promising to sense the surrounding environment. Being able to communicate and sense the vital sign of humans present in the environment will enable new vertical services of telecommunication, i.e., remote health monitoring. The proposed processing pipeline leverages spatially orthogonal beams to estimate the vital sign - breath rate and heart rate - of single and multiple persons in static scenarios from the raw Channel State Information samples. We consider both monostatic and bistatic sensing scenarios. For monostatic scenario, we employ the phase time-frequency calibration and Discrete Wavelet Transform to improve the performance compared to the conventional Fast Fourier Transform based methods. For bistatic scenario, we use K-means clustering algorithm to extract multi-person vital signs due to the distinct frequency-domain signal feature between single and multi-person scenarios. The results show that the estimated breath rate and heart rate reach below 2 beats per minute (bpm) error compared to the reference captured by on-body sensor for the single-person monostatic sensing scenario with body-transceiver distance up to 2 m, and the two-person bistatic sensing scenario with BS-UE distance up to 4 m. The presented work does not optimize the OFDM waveform parameters for sensing; it demonstrates a promising JCAS proof-of-concept in contact-free vital sign monitoring using mmWave multi-beam communication systems.

\end{abstract}

\begin{IEEEkeywords}
Joint communication and sensing, integrated sensing and communication, vital sign monitoring, sensing pipeline, mmWave multibeam communication 
\end{IEEEkeywords}

\section{Introduction}
With current mobile communication networks mature, the new era of mobile radio technologies will be far beyond communication alone. It is envisioned that the future generation of mobile communication, 6G and beyond, will offer truly intelligent wireless networks that provide both ubiquitous communication and high-accuracy localization and sensing services~\cite{6G_WhitePaper_Yang, 10005804,https://doi.org/10.48550/arxiv.2209.08847,9743500, 9924658, 10001163, 10000833}. 
Integrated sensing and communication (ISAC), or joint communication and sensing (JCAS), emerges as a decisive research topic to support not only high-speed communications but also novel vertical services such as autonomous driving and remote health monitoring. \par

\begin{figure}
    \centering
    \includegraphics[width=0.37\textwidth]{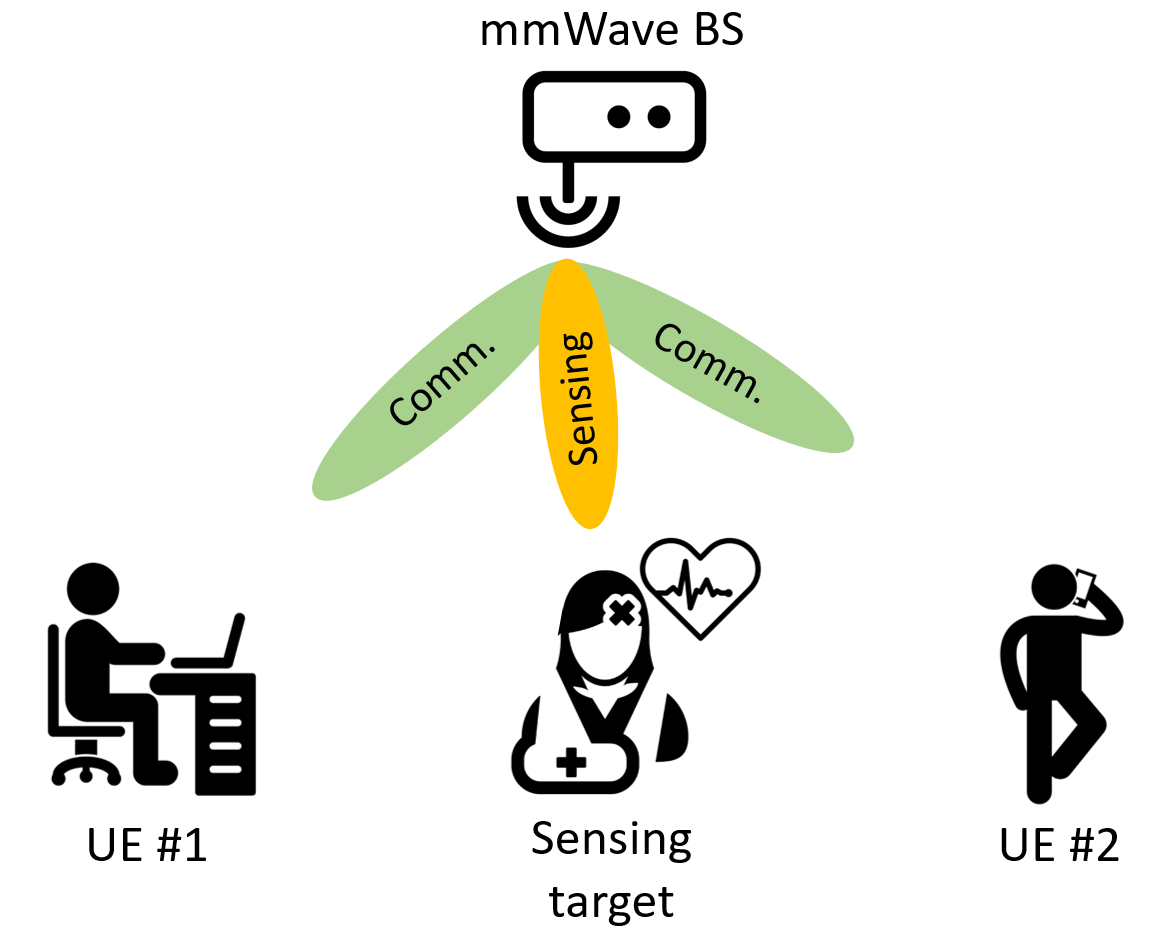}
    \caption{\centering{Illustration of JCAS systems with multi-beam BS 
    serving multiple UEs while sensing the
    targets} }
    \label{fig:jcas_intro}
\end{figure}

Multi-beam millimeter-wave (mmWave) radio systems take advantage of the spatial diversity of orthogonal beams, making it promising for simultaneous multi-user high-speed communication and accurate sensing~\cite{jcas1,jcas2,jcas4}. Orthogonal Frequency-Division Multiplexing (OFDM) waveform is employed in Long Term Evolution (LTE) and 5G New Radio (NR) due to its robustness against multipath fading, adaptive modulation and coding across subcarriers, as well as high flexibility in radio system design and radio resource management. OFDM is also promising for radar sensing, even with various symbols in the LTE and NR radio frames~\cite{8805161}. A multi-beam OFDM system transmits signals in orthogonal directions; the signals can be either scattered back from real-world targets and received at the Base Station (BS) for sensing or be propagated through the environment and arrive at the User Equipment (UE) for communicating. An illustration of a multi-beam JCAS system is depicted in Fig.~\ref{fig:jcas_intro}. 

Among the emerging vertical services of telecommunication, JCAS is one key enabler for remote health monitoring use cases.  
There are four main vital signs that can indicate a medical problem: body temperature, blood pressure, breath rate and heart rate; and we are interested in measuring the last two vital signs through sensing. Breathing is reflected in the chest movement and thus can be measured in the far field with an environment-embedded radio device. One complete breath includes both inhalation and exhalation, and the respiration/breathing rate is defined as the number of breaths completed per minute. The average adult’s respiration rate to heart rate ratio is approximately 1:4 when not exercising~\cite{scholkmann2019pulse}. 
Assuming this average relation, we could also extract heart rate from the sensing system. 
Being able to estimate vital signs and monitor health conditions 
is crucial for early disease diagnosis and prevention~\cite{intro1,intro2}. 

Current vital sign monitoring systems are predominated by wearable, contacting or intrusive solutions~\cite{intro3}. These techniques are either cumbersome, uncomfortable or not sustainable for long-term monitoring since these techniques typically require the target to initiate the monitoring process, which might lead to the risk of delayed medical issue indication and late medical treatment. 
Contact-free vital sign monitoring embeds the device in the environment, and thus, the target does not need to wear or carry any device in order to be sensed. This way, the sensing can be performed continuously without interrupting the target's daily activities~\cite{intro4}. However, the contact-free vital sign monitoring by environment-embedded radio devices faces some challenges: how to configure and deploy radio devices in the environment in order to measure and estimate vital signs accurately regardless of human orientation and distance to the device, how to differentiate vital signs from different persons present in the same environment and how to relate the measured signals to each target.

\subsection{State-of-the-art of contact-free vital sign estimation}
\label{sec:stateoftheart}

Researchers have utilized specific radar systems (e.g., ultra-wideband and mmWave MIMO~\cite{tiradar}) or communication systems with multi-nodes that is redundant for communication purpose (e.g., multiple distributed Wi-Fi modules) that can provide high-accuracy localization and real-time vital sign estimation in single and multi-person scenarios~\cite{intro5, Nature_paper, survey_VS}. 

Early work on sub-6~GHz radio-based vital sign monitoring can be found in~\cite{breathfinding, wifi1, wifi2}. These efforts focused on the Received Signal Strength (RSS) and the Channel State Information (CSI) phase acquired from WiFi systems to extract the breathing information of a single person. However, these efforts have practical limitations: 1) difficulties in recovering the reflected signals and separating the vital signs of different targets with the same breath rates due to the omni-directional nature of 2.4/5 GHz WiFi signals and the resulting multi-scattering between targets and objects; 2) lacking heartbeat estimation because the heartbeat movement is camouflaged by the larger displacement of the chest while breathing. 

Further WiFi vital signs estimation research in~\cite{phasebeat, breathtrack} proposed PhaseBeat, a method to exploit CSI phase difference data between receive antennas based on rigorous analysis with respect to its stability and periodicity. Initially, the CSI difference data is calibrated by removing direct current and high-frequency noises. After downsampling the cleaned data, the subcarrier with a larger variance for breath- and heart-rate estimation is selected. To resolve the multi-person scenario, a Root-MUSIC method is used to distinguish the different frequency tones, whereas, for the single-person scenario, an FFT-based peak search is performed to extract the estimated rates. The estimation accuracy of breath- and heart rates achieves $>$95\%, 95\%, 90\% and 80\% for 1, 2, 3 and 4-person scenarios, respectively, with respect to ground-truth obtained with a NEULOG Respiration Monitor and a fingertip pulse oximeter. Moreover, it is shown that estimation performance degrades as the distance between the transmitter (Tx) and receiver (RX) increases. At the distance of 1~m, 0.15 beats per minute (bpm) mean estimation error can be obtained, whereas the mean estimation error increases to 0.2~bpm and above 0.4~bpm at the distance of 5~m and 11~m, respectively. 

In~\cite{breathtrack}, a joint Angle of Arrival - Time of Flight (AoA-ToF) beamformer is proposed at each packet time. To deal with the small ToF resolution in 40 MHz commodity WiFi systems, a Singular Value Decomposition (SVD) algorithm is utilized to reformulate the problem with reduced dimensions by decomposing the data matrix into the signal and noise subspaces. The estimated AoA-ToF grid after SVD is then multiplied by the time domain symbol signal, which yields a time domain signal representing the phase variations due to chest displacement. The single-person scenario in line-of-sight (LoS) and non-light-of-sight (NLoS) conditions with up to 4 m Tx-Rx distance yields $>$99\% median accuracy.

In~\cite{holo}, \textit{Eid et al.} propose HoloTag, an ultra low-cost Ultra High Frequency (UHF) Radio-Frequency Identification (RFID) array-based system over which a holographic projection of its environment is measured and utilized to localize and monitor the vital signs of several targets. Using Minimum Variance 
Distortionless Response (MVDR) beamforming for AoA estimation and an FFT-based vital sign estimation method, the proposed system achieves 11\textdegree~and 17\textdegree~AoA median error for a single and two-person scenario, respectively. The breathing rate estimation error is kept under 0.4 bpm and 0.5 bpm for more than 60\% trials in single-person and two-person scenarios, respectively. 

\textit{Zhang et al.} presented MTrack \cite{mtrack}, a 2~GHz bandwidth system that generates signals between 4-6~GHz. Equipped with one Tx antenna and 16 Rx antennas, the system relies on AoA-ToF beamforming with a path selection algorithm to suppress interference from dynamic multi-paths in order to detect and track individual signals under multi-person conditions. The proposed system is capable of localizing and tracking people's trajectories with a 20~cm error at a maximum distance of 7 m for both LoS and NLoS scenarios. Vital sign estimation achieves $>$99\% median accuracy for both breath- and heartbeat rates. \textit{Chenglong et al.}\cite{chenglonglocalization} establish a sub-6 GHz radar-like MIMO-OFDM prototype for contact-free localization and human tracking in real-time. In~\cite{radar8}, an 8 GHz FMCW radar is used alongside a multi-person tracking algorithm to localize targets and extract their time domain phase difference data for estimating chest displacements due to breathing and heartbeat activities. From the presented experiments, the algorithm is capable of keeping the error below 3 bpm during 90\% of the measured time.

At mmWave frequencies, research efforts are mainly focused on the use of radar systems. In~\cite{radar771, radar77}, a 77 GHz FMCW MIMO radar is used to localize humans in the range and angular domains before extracting their time domain phase difference data. Evaluations under a two-person scenario show that the system is capable of achieving errors of less than 1 bpm breath rate, and 3 bpm heart rate at a target-to-radar distance of 1.6 m, with a minimal 40\textdegree~angular separation between targets. In the 60~GHz band, \cite{radar60} utilizes maximum ratio combining (MRC) on a UWB 2$\times$4 MIMO radar to improve the signal-to-noise ratio (SNR). This technique leads to an 18\% improvement in heart rate estimation accuracy with respect to estimation obtained directly from each Tx-Rx branch. Another work in~\cite{devalg} uses the 24 GHz FMCW radar to capture time-domain phase signals for breath- and heart rate estimations in the single-person scenario. This includes detrending the original signal and performing continuous wavelet transform (CWT) to obtain fine-tuned frequency subspaces. CWT is also used in~\cite{radar8} to separate breath- and heart rate frequencies. In \cite{alidoustaghdam2023enhancing}, the authors have utilized human pose estimation to find the chest location and then beamforming towards the chest to estimate the vital signs better.

\subsection{Contributions of this work}
\label{subsec:goal}


To the best of our knowledge, there is no experimental investigation on leveraging the mmWave OFDM communication devices with standard communication waveforms for vital sign estimation. Without doing so, we can not evaluate whether mmWave JCAS using OFDM waveforms for communication and sensing can be a reliable solution for remote health monitoring. 
The goal of this paper is thus to leverage our 26 GHz multi-beam OFDM communication systems with waveform profiles following the LTE standard (transferable to 5G NR and beyond) for proof-of-concept investigations on the contact-free vital sign sensing of up to two persons present in the same environment. 
{In our envisioned research, we follow a step-wise approach to deal with realistic dynamic scenarios: 1) localization when a human moving in space; 2) pose estimation when a human has limb movement; 3) vital sign estimation focusing on human chest. The scope of this paper lies in 3) when a human has no limb movement and is static in space. Even though only focusing on 3) in this paper, the proposed approach is highly likely to be able to be used in practical dynamic scenarios, after localization and pose estimation.}

Our transceiver (TRx) setup is considered to mimic BS monostatic sensing and BS-UE bistatic sensing scenarios. Despite using the static target(s), both the frontal and side orientations of humans relative to the device are investigated. We exploit the spatial diversity in multi-beam transmission along with frequency diversity in the OFDM sub-carriers to obtain vital sign information. The main contributions of this paper are the proof-of-concept measurement campaign, the proposed CSI pre-processing and breath/heart rate estimation pipelines, and the performance evaluation through extensive measurements.




\section{System and Experiment Overview}
\label{sec:system_experiment}

\subsection{Prototype and experiment}
\label{sec:testbed}

\begin{figure}[t]
    \centering
    \includegraphics[width= 0.48\textwidth]
{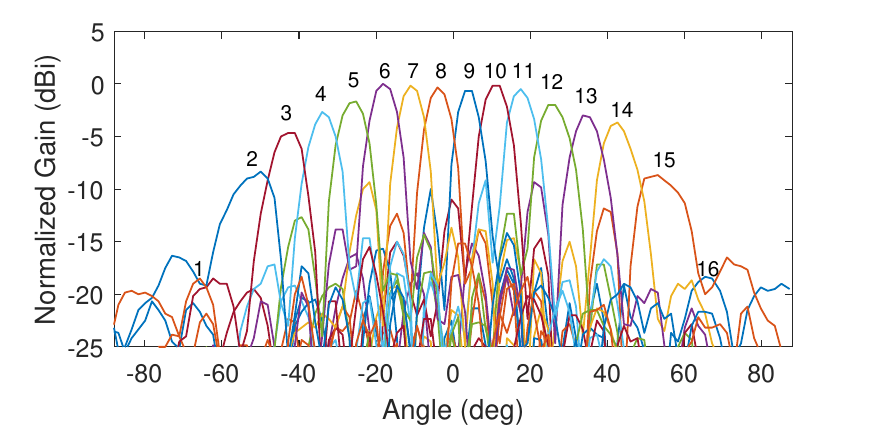}
    \caption{\centering{Beam pattern of the Butler matrix at 26 GHz}}
    \label{fig:beampattern}
\end{figure}

The KU Leuven mmWave MIMO testbed 
consists of a pair of Butler matrix units, a mmWave front-end operating at the $26$~GHz band and multiple Universal Software Radio Peripheral (USRP) devices. The Butler matrix used in this testbed is the one thoroughly depicted in~\cite{testbed1}. The system is able to either transmit or receive up to 16 frequency dependent spatially-orthogonal beams. The beam pattern of the Butler matrix at 26 GHz is shown in Fig.~\ref{fig:beampattern}. It is noticeable how the center beams have narrower beamwidth and higher gain, up to 20 dB, compared to those at the edge. 
For the bistatic investigation in this paper, only the uplink BS-UE transmission is considered, namely the UE acts as the transmitter and the BS is the receiver. 

The USRP transmitting intermediate frequency $f_{IF}$ of 2.4 GHz with 20 MHz bandwidth is used to generate the $f_{RF}$ mmWave signal in the transmitter end. The IF signal serves as input to the Butler matrix, and the ERASynth+ RF signal generator with the operating frequency of $f_{LO} = 11.8$ GHz is connected to the local oscillator port of the Butler matrix~\cite{testbed4}. The RF up/down conversion in a Butler matrix follows the relation of $f_{RF} = 2f_{LO}+f_{IF}$~\cite{testbed2}. In the receiver end of the Butler matrix, all 16 beams are used to receive signals from different directions. Therefore, all available ports of the Rx Butler matrix are employed and connected to the 16 input ports of the receiver USRPs, while the local oscillator port of the Butler matrix is connected to the signal generator, which is synchronized for both Tx and Rx Butler matrices using Pulse Per Second (PPS) and 10 MHz clock reference signals. 

All USRPs run LabView Communications MIMO Application Framework~\cite{testbed3} with a Time Division Duplexing (TDD) frame structure with OFDM symbols. 
This communication system has a transmit power $P_T$ of 15 dBm, and a 31 dB gain is introduced at the receiver side. 
The symbol duration is 66.67 $\mu$s (consistent with LTE and 5G NR standards) with a cyclic prefix duration of 5.21 $\mu$s. We transmit 1 uplink pilot symbol per subframe slot\footnote{Despite that we use one UE in our investigation in this paper, the available testbed is capable of multi-user communication. There are 1200 subcarriers (with 15 kHz spacing for 20 MHz bandwidth, specified by 3GPP for LTE standard) in use, which are divided into 100 resource blocks to support up to 12 users. Each UE transmits a pilot on a different subcarrier in such a single resource block. The pilot tone of each UE consists of 100 sub-carriers, which are evenly spaced in a frequency band of 20 MHz. The BS receives the orthogonal pilots sent by the UEs simultaneously and distinguishes between the different UEs by the frequency interleaving of the subcarriers. In this way, the channel is captured between the UEs and the BS antennas for 100 subcarriers. }. The slot has a duration of 0.5 ms consisting of 7 OFDM symbols and cyclic prefixes. 
The complete radio frame format used in the testbed can be found in~\cite{ni}. The modulation scheme used is 16-QAM. The system parameters setting are summarized in Table~\ref{tab:par}.

\begin{table}[t]
\centering
\captionsetup{justification=centering, labelsep=newline}
\caption{{mmWave Multi-beam testbed parameters setting}}
\label{tab:par}
\resizebox{0.25\textwidth}{!}{%
\begin{tabular}{lc}
\hline
\textbf{System parameter} & \textbf{Value} \\ \hline
Transmit power            & 15 dBm         \\
Rx Gain                   & 31 dB          \\
Center frequency          & 26 GHz         \\
Bandwidth                 & 20 MHz         \\
Subcarrier spacing        & 15 kHz         \\
Subcarriers               & 100            \\
Modulation                & 16-QAM            \\
Sampling rate             & 30.72 MHz         \\ \hline
\end{tabular}}
\end{table}


\begin{figure}[t]
\begin{subfigure}[t]{0.33\linewidth} 
    \centering
    \includegraphics[width = \linewidth]{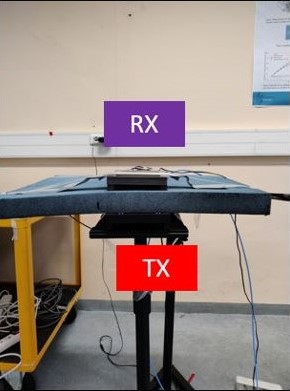}
    \caption{\centering{Mono-static}}
    \label{fig:meas1_a}
  \end{subfigure}%
  \hspace{0.01em}
   \begin{subfigure}[t]{0.655\linewidth} 
    \centering
    \includegraphics[width = \linewidth]{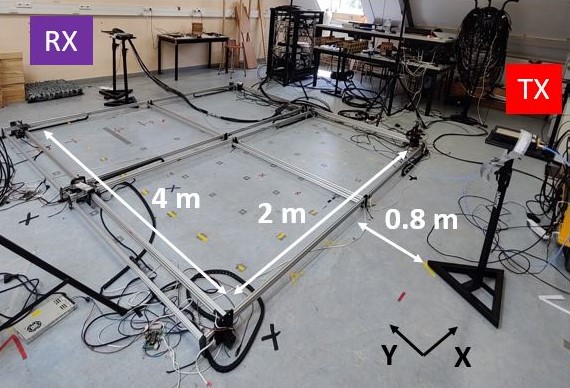}
    \caption{\centering{Bistatic}}
    \label{fig:meas1_b}
  \end{subfigure}
    \caption{\centering{Tx and Rx measurement setup}}
    \label{fig:meas1}
\end{figure}  


\begin{figure*}[t]
    \centering
      \begin{subfigure}[b]{0.28\textwidth}
         \centering
         \includegraphics[width=\textwidth]{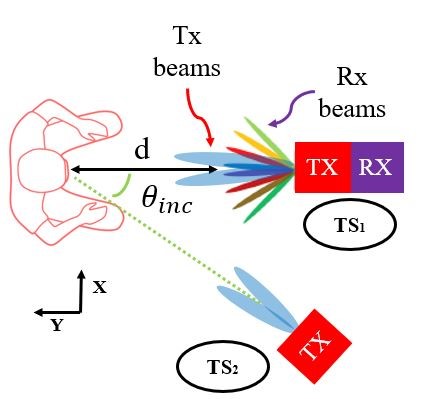}
         \caption{$\text{TS}_1$ and $\text{TS}_2$}
         \label{fig:ts_1_2}
     \end{subfigure}%
     \hspace{0.6cm}
     \begin{subfigure}[b]{0.45\textwidth}
         \centering
         \includegraphics[width=\textwidth]{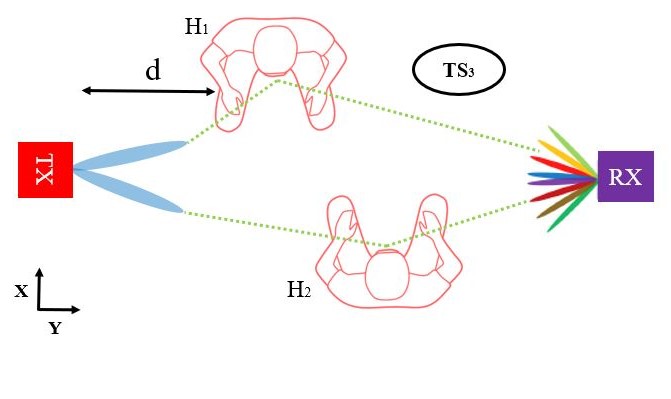}
         \caption{$\text{TS}_3$}
         \label{fig:ts_3}
     \end{subfigure}

    \caption{\centering{Testing scenarios. $\text{TS}_1$: single-person with mono-static TRx configuration. $\text{TS}_2$: single-person with bistatic TRx configuration (varying incidence angles). $\text{TS}_3$: two persons with bistatic TRx configuration.}}
    \label{fig:tsscheme}
\end{figure*}

Both monostatic and bistatic sensing scenarios are considered by placing the Tx and Rx as shown in Fig.~\ref{fig:meas1}. The measurement took place in an indoor lab environment with the size of $11 \times 6.5$ m$^2$ 
Three testing scenarios (TS) were performed to obtain measurement data conveying different perspectives for vital sign estimation. 
On the Tx side, only the 2 central beams (beam 8 and 9) 
were employed\footnote{{
The main reason behind utilizing only 2 out of 16 available Tx beams is to illuminate up to two targets in our scenario, aiming to distinguish between these targets. While it's technically feasible to connect multiple USRPs to the Tx Butler matrix for more Tx beams and detect more than two targets using orthogonal pilots, using additional beams not directed at human chests is unnecessary for breathing/heartbeat rate estimation and leads to unnecessary post-processing. We specifically use 2 beams at the center (beam 8 and 9) due to their higher beamforming gain compared to other beams.}},  
while all 16 beams were used on the Rx side. We consider uplink transmission, thus the UE and BS become the Tx and the Rx, respectively. 
In all experiments, 5 seconds of data are captured, corresponding to a total of 10000 symbols. 
To capture the ground truth of the heart- and breath rate, we used the TMSi Mobi devices~\cite{mobi} in which the heart rate is measured via a pulse oximeter fingertip sensor and the breathing rate is obtained via a respiratory monitor belt around the chest to measure its displacement.

A normal adult's breathing rate is about 12 - 16 per minute (from 1 to 1.23 per 5 seconds), and the heart rate is ranging from 60 to 100 beats per minute (from 5 to 8.3 per 5 seconds). 
The reason to limit the measurement time up to 5 seconds is mainly due to the data size of 10000 symbols of raw I/Q samples of 100 subcarriers and $16 \times 2$ beam pairs as well as the resulting processing pressure of them. {Furthermore, previous work in \cite{10081872} demonstrates that 5 seconds of observation is acceptable for real-time vital sign monitoring.
In practice, from the hardware perspective, there is a data buffer storage limit for storing and processing the real-time data.
}
In addition, in a joint communication and sensing context, one would expect that the infrastructure can estimate the vital signs as real-time as possible, i.e., by using a few symbols, so that predictive measures could be taken immediately. 

\subsection{Testing scenarios}
We consider three testing scenarios (TS) representing both monostatic and bistatic setup configurations with up to two measurement subjects.   
\subsubsection{$\text{TS}_1$}
\label{sec:ts1}
The first scenario utilizes the monostatic (co-located) TRx setup as shown in Fig.~\ref{fig:meas1_a}. We placed an absorbing foambetween Tx and Rx to suppress the unwanted cross-talk between each other\footnote{
{In practical engineering for experimental scenarios, at least two aspects can mitigate crosstalk: physical isolation of transmit and receive antennas (such as increasing the distance between them, altering their beam directions, or using orthogonal polarization) and employing signal processing techniques like interference cancellation (e.g., filtering).}
}. We positioned a human target sitting on a chair in front of the TRx module with the distance between the human and the device varied from 1~m up to 2~m, as depicted in Fig.~\ref{fig:ts_1_2}, to study the effect of distance on the vital sign estimation using monostatic setup. 

\subsubsection{$\text{TS}_2$}
\label{sec:ts2}
The second scenario aims to study the influence of the incidence angle of Tx beams on the vital sign estimation in a bistatic TRx setup, as shown in Fig.~\ref{fig:meas1_b}. Thanks to the multiple beams employed at Rx, 
accurate estimates can be obtained 
based on the multi-beam spatial diversity, 
regardless of the relative orientations between the body and the incidence beam. 
A schematic diagram of $\text{TS}_2$ is shown in Fig.~\ref{fig:ts_1_2}, where different incident angles $\theta_{inc}$ were tested by changing the Tx location. 
The Rx module is placed in front of the human target while the distance between the human and the Rx is varied from 1 to 2 m for each Tx location. For the Tx location where the cross-range between Tx and Rx in view of target is 0.95 m, the incidence angles $\theta_{inc}$ are 25.4\textdegree, 32.5\textdegree~and 43.5\textdegree~for distances between target and Rx of 2~m, 1.5~m and 1~m, respectively. Note that $\theta_{inc}$ in $\text{TS}_1$ is 0\textdegree. 

Given a human shoulder width of $W$~m, with the target-to-Rx distance of $d$~m, the required Rx azimuth Field of View (FoV) to cover such width 
can be expressed as:
\begin{equation}
\label{eq:fov}
    \Delta_{FoV} = 2\arctan{\left( \frac{W}{2d} \right)}.
\end{equation}
In both $\text{TS}_1$ and $\text{TS}_2$ scenarios, the human target performed 
3 breath cycles during 5~s period. The exact breathing and heart rate were also captured using the ground-truth sensor. 

\subsubsection{$\text{TS}_3$}
\label{sec:ts3}
The third scenario involves two persons in a bistatic setup, as shown in Fig.~\ref{fig:ts_3}. The distance between Tx and Rx is 4 m, where the Tx and Rx face each other mimicking the real-world mmWave communication scenario. Two human targets were placed inside the FoV of the Rx beams and the Fresnel zone of the Tx and Rx. The distance between target 1, denoted as $\text{H}_1$, and the Tx was varied from 1 to 3~m in the $Y$ axis, while target 2, denoted as $\text{H}_2$, was always placed at 2~m away from the Tx in the $Y$ axis. The separation between targets in the $X$ axis was kept to 1~m. For all measurements, $\text{H}_1$ had a breath rate lower than that of $\text{H}_2$. Both tried to keep their breath constant during the measurement. Since only one ground-truth sensor was available, simultaneous ground-truth measurement on both targets was not possible; measurements were repeated with the Mobi device on each target respectively and ground-truth values were obtained as the average rates of all measurements. This hardware limitation needs to be considered when comparing the estimation results with the ground-truth values.

\section{Vital Sign Estimation Pipeline}

\label{sec:meascamp} 
\begin{figure*}[t]
    \centering
       \includegraphics[width=0.95\textwidth]{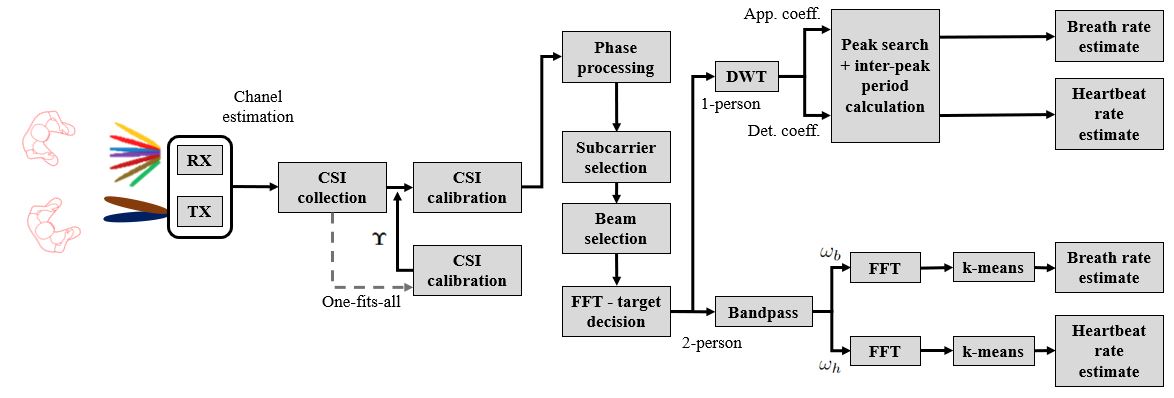}
    \caption{Data processing pipeline for the breathing- and heart rate estimation.}
    \label{fig:pipeline}
\end{figure*}

\label{sec:method}

This section outlines a processing pipeline for directly estimating breath and heart rates from raw CSI samples in a multi-beam mmWave communication testbed. The pipeline addresses time-frequency calibration to reduce the impact of frequency up/down conversion imperfections on measured CSI amplitude and phase. It distinguishes between estimation methods for single and multi-person scenarios, with the latter requiring clustering. Furthermore, it highlights the advantage of a multi-beam system for vital sign estimation, even when the OFDM parameters in the communication standard may not be ideal for sensing.

\subsection{Proposed pipeline: overview}

\label{sec:pipelineoverview}

The proposed data processing pipeline for the breath- and heart rate estimation is depicted in Fig.~\ref{fig:pipeline}. For a given subcarrier and symbol, the Channel Transfer Function (CTF) is estimated in the uplink and is denoted as $\bm{H}\in  \mathbb{C}^{N_R \times N_T}$:
\begin{equation}
    \bm{H}(f,t_s) =  \begin{bmatrix}
h_{1,1}(f,t_s) & . & . & . & h_{1,N_T}(f,t_s) \\
. & . &  &  & . \\
. &  & . &  & . \\
. &  &  & . & . \\
h_{N_R,1}(f,t_s) & . & . & . & h_{N_R,N_T}(f,t_s) 
\end{bmatrix} ,
\end{equation}
where $h_{n_r,n_t}$ is the matrix element of $\bm{H}(f, t_s)$ at each sampling frequency $f$ and symbol $t_s$. $N_R$ and $N_T$ are the total numbers of the Rx and the Tx beams, respectively. $n_r = 1, ..., 16$ and $n_t = 1,2$ are the indices of Rx and Tx beams. 
In total, $N_{f}\times N_s$ are captured in $\bm{H}$, where $N_{f}$ and $N_{s}$ denote the total number of transmitted symbols and subcarriers, respectively. 

When a transmitted signal is reflected or back-scattered by the human chest with a  breathing pattern frequency of $f'_b$ and a heart rate frequency $f'_h$, the phase of the reflected signal is also periodic with the same breathing and heartbeat frequencies~\cite{mtrack,csi1}. 
In the far-field scenario, the transmitted signal is back-scattered from the human chest as a plane wave. 
For a given transmit and receive beam-pair with $n_f= 1, 2, \dots, N_f$ subcarriers and $n_s = 1,2, \dots, N_s$ symbols, the can be expressed as~\cite{mtrack,csi1}:
\begin{equation}
\label{eq:csi}
    \angle H_{nr,nt} \left( f(n_f),t_s(n_s) \right) = 
    2\pi\frac{d(t)}{\lambda_{n_f}},
\end{equation}
where $H_{nr,nt}$ is one element of matrix $\bm{H}$ and $d(t) = D + \alpha_b \cos{(2\pi f'_bt)}+\alpha_h \cos{(2\pi f'_ht)}$ is the propagation distance of the back-scattered signal influenced by the periodic rise and fall of the human chest. When the human stands still, $D$ is a constant and is the mean distance of the back-scattered path. $\alpha_b$ and $\alpha_h$ are the corresponding amplitudes of the periodic signal from chest movements due to breath and heartbeat activities, respectively. $\lambda_{n_f}$ is the wavelength of subcarrier $f(n_f)$. Eq.~\eqref{eq:csi} shows that the measured (unwrapped) phase increases as the subcarrier frequency increases. The received time domain phase information is 
affected by the breathing and heartbeat activities of the human body, and contains the necessary information for vital signs estimation. Therefore, the calibration and denoising of the measured signal before obtaining relevant estimates are crucial steps, as discussed in~Sec.~\ref{sec:calibration}.

Following calibration and pre-processing of the measured channel data, power analysis is conducted to select appropriate beams that convey information about human vital signs, filtering out beams that do not contain such information. Additionally, subcarrier selection is employed to capture phase information from subcarriers with higher time domain variance, as elaborated in~Sec.~\ref{sec:vital_sign_est}. 


In the single-person scenario, we utilize the Discrete Wavelet Transform (DWT) on the calibrated and pre-processed channel data to focus on the relevant frequency bands associated with breathing and heartbeat. DWT offers a multi-scale, multi-resolution analysis of the time-frequency representation, as opposed to FFT analysis. 
By combining DWT and peak search, the estimated breath- and heart rates are obtained by taking the average of all the inter-peak interval periods from the low and high-frequency reconstructed signals, respectively. 

In the multi-person scenario, the inter-peak interval calculation used in the single-person scenario can not be used to compute more than one period rate. Moreover, as the discrepancies in breath and/or heartbeat rates between the two persons are not known in practice, we can not use the DWT to decompose the signal into the frequency bands related to each of the targets' vital signs activity. An alternative solution to multi-person estimation could be to apply an FFT to the decomposed signal after the DWT to obtain the frequency domain behavior of the phase data. Nevertheless, we discovered that the iterative DWT decomposition steps deteriorate the already-weak phase periodic signal, hence failing to capture the different frequency tones for multiple human targets. Therefore, our proposed pipeline employs bandpass filters to segregate breathing and heartbeat rates into their respective frequency bands in multi-person scenarios. Referring to the method used in mmWave FMCW radar (as described in~\cite{radar771}), we employ frequency domain analysis and perform an FFT peak search to estimate the number of persons in the measured scenario. 
We apply the estimated number of humans as input for clustering in the $k$-means algorithm to identify multiple frequency tones in both breathing and heartbeat frequency bands.  

\subsection{Calibration and pre-processing of measured data}
\label{sec:calibration}

During channel estimation, the measured raw CSI suffers from various frequency-dependent phase errors due to imperfect synchronization between the transmitter and the receiver  \cite{datacal1}. In addition, the estimated CSI phases are influenced by Sampling Frequency Offset (SFO) and Symbol Timing Offset (STO)~\cite{datacal2}. The implementation of OFDM is susceptible to the effect of In-phase and Quadrature-phase (IQ) imbalance in analog processing, leading to non-linear phase distortion in CSI estimation. Moreover, a random initial phase generated by the local oscillator and the consequent imperfect compensation of the phase-locked loop may introduce Carrier Phase Offset (CPO) in the received phases~\cite{datacal3}. 

In our previous work~\cite{datacal1}, a method to compensate for the introduced errors in the phase of the measured channel data has been presented. The estimated CSI $\hat{H}_{n_r,n_t}$ at a given symbol and subcarrier in the presence of the aforementioned errors can deviate from the truth $H_{n_r,n_t}$ in the way as:
\begin{equation}
\label{eq:chanelcal}
\hat{H}_{n_r,n_t} = H_{n_r,n_t} \exp{\left( -j \left( \zeta_{SFO}+\zeta_{STO}+\zeta_{IQ}+\eta_{CPO}\right) \right)},
\end{equation}
where $\zeta_{SFO}$, $\zeta_{STO}$ and $\zeta_{IQ}$ are the phase shift caused by SFO, STO, and IQ imbalance, respectively. The SFO phase shift is proportional to the subcarrier index, and the $\zeta_{IQ}$ for a given subcarrier $f(n_f)$ is given by:
\begin{equation}
    \zeta^{(n_f)}_{IQ} = \arctan{\left( \epsilon_g \frac{\sin{(n_f \xi_{t}+\epsilon_p)}}{\cos{(n_f\xi_t)}} \right )},
\end{equation}
where $\epsilon_g$, $\epsilon_p$ represent the gain and phase mismatch, and $\xi_t$ is the unknown time offset, respectively. Finally, $\eta_{CPO}$ can be considered as a random constant after the initiation of the transmitter. The aforementioned phase shifts errors are calibrated by the following nonlinear regression along the subcarriers:
\begin{equation}
    \arg \min_{\bm{\Upsilon}} \sum_{n_f}\left( \Delta \Theta_{n_f}-\zeta_{IQ}^{(n_f)}-n_f  \zeta_{SFO/STO}-\eta_{CPO} \right)^2,\
\end{equation}
where $\Delta \Theta_{n_f}$ is the measured residual phase at subcarrier $n_f$, and $\bm{\Upsilon} = [\epsilon_g, \epsilon_p, \xi_t, \zeta_{SFO/STO}, \eta_{CPO}]$. Such problem can be solved using the Levenberg-Marquardt algorithm~\cite{datacal1}. 

Once the phase errors are estimated, we obtain the calibrated channel by applying Eq.~\eqref{eq:chanelcal} to compensate for these errors. 
Note that for a given measurement configuration, these parameters only need to be estimated once, and can be used for future CSI calibration without the redundant processing, effectively adopting a one-fits-all calibration approach~\cite{chenglonglocalization}. Additionally, we apply MATLAB's \textit{rloess} smooth filter on the calibrated data to mitigate rapid fast phase variations~\cite{smooth} that may occur in the data post-calibration. This filter provides robust local regression using weighted linear least squares and a 2nd-degree polynomial model.


To assess the performance of our calibration method, we configured the Tx and Rx as collocated as in $\text{TS}_1$. To replicate the chest's rise and fall, a person holding a metallic board at chest height is positioned at a distance of 1 meter from the TRx. The periodicity of the backscattered signal from the metallic board is expected to be significantly more pronounced than that from a human body. In this testing scenario, 
four complete rise-fall cycles were performed to mimic breathing within a 5-second interval. We utilize both central beams for transmission and all 16 beams for reception.

\begin{figure}[tb]
     \centering
     \begin{subfigure}[b]{0.24\textwidth}
         \centering
         \includegraphics[width=\textwidth]{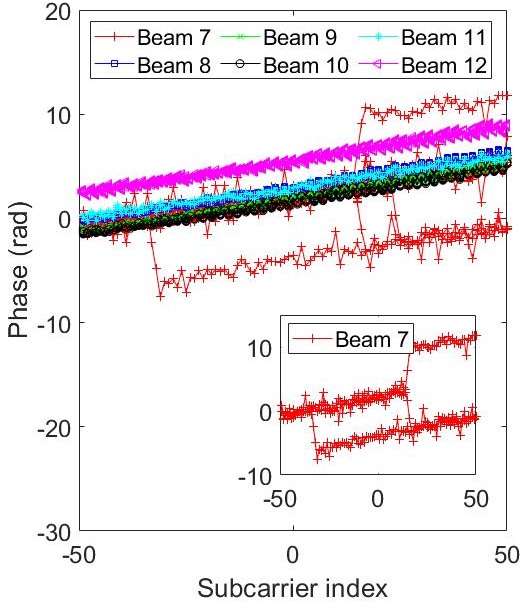}
         \caption{Raw phase}
         \label{fig:datacalraw}
     \end{subfigure}
     \hfill
     \begin{subfigure}[b]{0.24\textwidth}
         \centering
         \includegraphics[width=\textwidth]{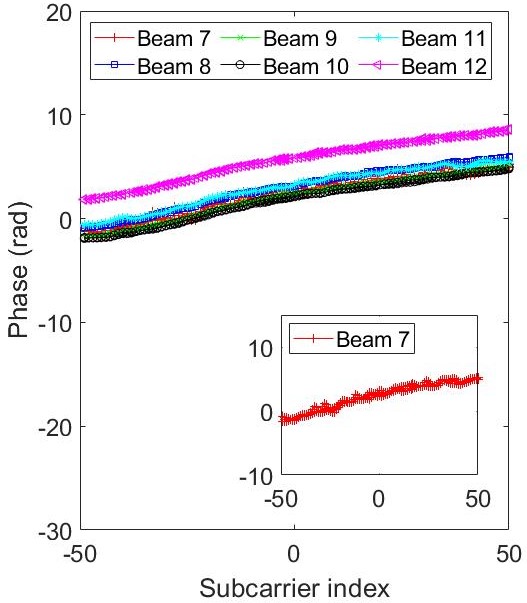}
         \caption{Calibrated phase}
         \label{fig:datacalcal}
     \end{subfigure}
     \caption{\centering{Raw vs calibrated CSI phase for the Tx beam 9 and the Rx beams covering the human over 5 transmitted symbols across all subcarriers (subcarrier index 0 indicating the central frequency)
     }}
     \label{fig:datacal}
\end{figure}

\begin{figure}[tb]
         \begin{subfigure}[b]{0.23\textwidth}
         \centering
         \includegraphics[width=\textwidth]{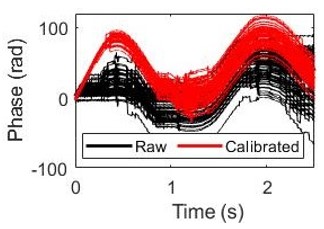}
         \caption{Tx beam 9 -- Rx beam 8}
         \label{fig:datacal3v1}
     \end{subfigure}
     \hfill
         \begin{subfigure}[b]{0.23\textwidth}
         \centering
         \includegraphics[width=\textwidth]{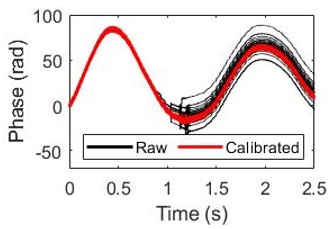}
         \caption{Tx beam 9 -- Rx beam 9}
         \label{fig:datacal3v2}
     \end{subfigure}
         \caption{\centering{CSI phase progression of 100 subcarriers over 2.5 s capture time 
         of a human with a metallic board mimicking a human's breathing cycle}}
    \label{fig:datcal3}
\end{figure}

Fig.~\ref{fig:datacal} presents a comparison between the raw channel phase and the calibrated phase for the beams covering the human subject. In general, the aforementioned frequency domain errors do not manifest uniformly across all beam pairs; instead, they tend to sporadically appear over certain beams during specific symbols.
Fig.~\ref{fig:datacalraw} indicates that the channel between Tx beam 9 and Rx beam 7 exhibits variations across subcarriers during multiple transmitted symbols, whereas the remaining 
beam pairs display a linear relationship with subcarrier increase, aligning with the expected behavior of CSI phase with respect to subcarrier in Eq.~\eqref{eq:csi}. 
Fig.~\ref{fig:datacalcal} demonstrates that the calibration implementation effectively rectifies phase errors across the frequency domain. The phase progression over the 2.5 s data capture is then recovered with the proposed calibration method, validating its effectiveness, 
as shown in In Fig.~\ref{fig:datcal3}. 

After applying phase calibration, we proceed with data pre-processing to eliminate undesirable DC and high-frequency noise components, which can negatively impact the accuracy of vital sign estimation for the following reasons:
1) the DC component of the signal affects the subcarrier selection and further damages estimation based on peak search in the frequency domain, as it appears as a large peak at $f = 0$ Hz; 2) high-frequency noises camouflage the actual frequency tones related to vital sign activity. 
Considering that the DC component is expected to exhibit minimal variation throughout the signal's duration, we employ 
a Hampel filter with a large sliding window size of of 2000 samples (an empirical value derived from our measurement data) and a small threshold of $0.01\sigma$, 
where $\sigma$ represents the standard deviation of the time domain phase samples. This filter captures the basic trend of the original data. Subsequently, we obtain the DC-removed phase difference data by subtracting the basic trend from the original data. 

\begin{figure*}[tbh]
    \centering
      \begin{subfigure}[t]{0.3\textwidth}
         \centering
         \includegraphics[width=\textwidth]{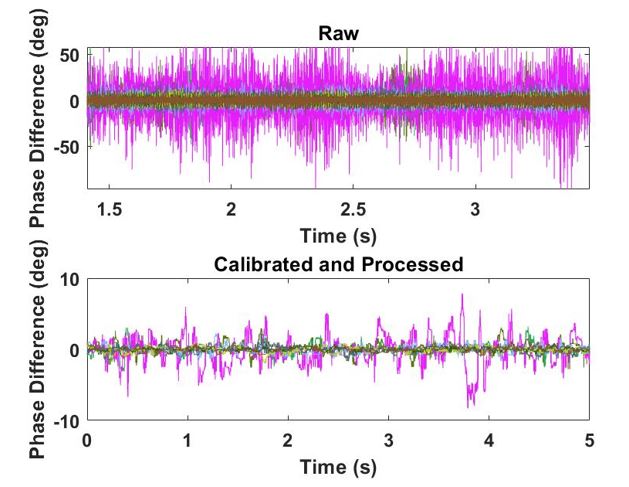}
         \caption{\centering{Raw phase difference data before and after the calibration and pre-processing with Hampel filter}}
         \label{fig:calproc1}
     \end{subfigure}%
     \hspace{0.1em}
     \begin{subfigure}[t]{0.3\textwidth}
         \centering
         \includegraphics[width=\textwidth]{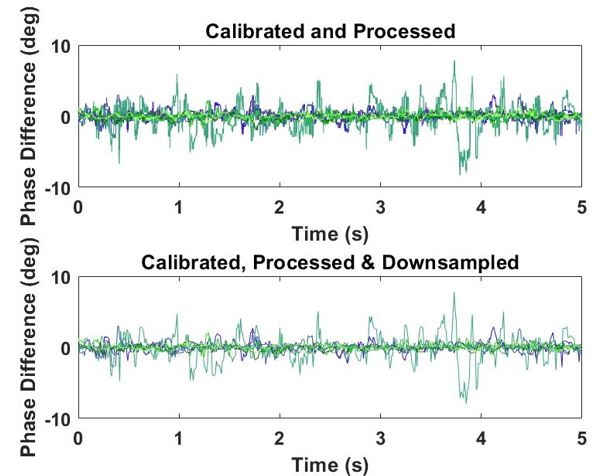}
         \caption{\centering{Effect of down-sampling on the calibrated and pre-processed dat}}
         \label{fig:calproc2}
     \end{subfigure}%
          \hspace{0.1em}
     \begin{subfigure}[t]{0.38\textwidth}
         \centering
         \includegraphics[width=\textwidth]{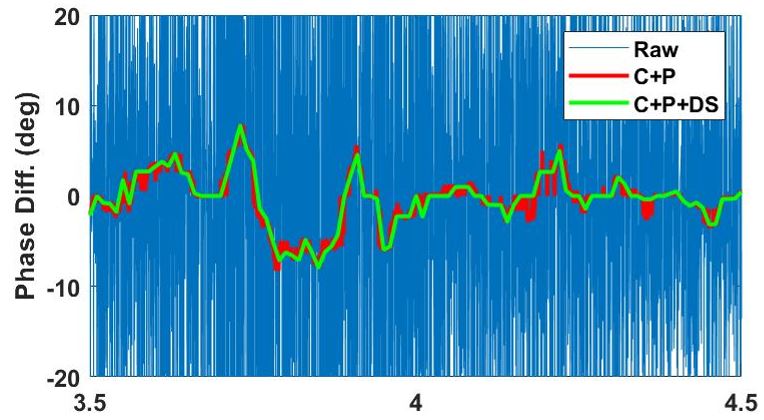}
         \caption{\centering{Time domain zoom-in of the calibrated, pre-processed and down-sampled data}}
         \label{fig:calproc3}
     \end{subfigure}

    \caption{Example of the calibrated and pre-processed data using the pair Tx Beam 9 -- Rx Beam 9
    }
    \label{fig:calproc}
\end{figure*}

To remove high-frequency noises, a Hampel filter with a reduced window size is necessary to filter out the rapid phase variations. 
Hence, we use a window size of 50 samples and a threshold of $0.01\sigma$. Given the relatively small phase variation attributed to vital signs activity, we found that a small threshold value is preferred, with $0.01\sigma$ being a value leading to proper removal of DC and high-frequency noises. Finally, down-sampling is performed to lower the original high sampling frequency. Fig.~\ref{fig:calproc} shows examples of the calibrated and pre-processed data.   

\subsection{Power analysis}

Our multi-beam testbed pertains to spatial diversity, where each individual receiving beam obtains information from a specific AoA range due to its narrow beamwidth. 
By multiplexing the FoV of the orthogonal multi-beam set-up, we can cover a large area of interest.
Beam selectivity needs to be implemented to make sure that the beams carrying information about the human target are selected and further processed, while other beams are filtered out. 
We mainly investigate the time domain amplitude variations of the measured CSI samples among the beams. 

To obtain the information of the back-scattered power from the human body in the delay ($\tau$) domain, we obtain the Power Delay Profile (PDP) for a given Tx--Rx beam pair link and a given symbol from the CTF as~\cite{testbed4}:
\begin{equation}
    \text{PDP}_{n_t,n_r}(:,t_s) = \left| \text{IFFT}\{\text{Hann}\{h_{n_t,n_r}(:,t_s)\}\} \right|^2,
\end{equation}
where a Hanning window is multiplied with the CTF to suppress side lobes. 
Considering the relatively limited bandwidth of our testbed, the power associated with the target becomes concentrated within the first PDP peak.

We analyse the reflection/back-scattering power loss on the human target. To do so, the effect of the distance-dependent path loss needs to be removed. From link budget theory, for a given transmission power $P_T$, equal transmit and receive antenna gains $G_T = G_R = G$ and received power $P_R$, the total loss $L = P_T+G_T+G_R-P_R$ is equivalent to:
\begin{equation}
    L = L_P(d_T)+L_P(d_R)+L_R,  
\end{equation}
where $L_P$ is the path loss in free-space, $L_R$ is the back-scattered loss, and $d_T$ and $d_R$ are the distances of the object away from Tx and Rx, respectively, where $d_T$ and $d_R$ are equal in the case of amonostatic radar.

We define the power back-scattering coefficient, $r_L$, as a metric to measure the reflection loss encountered by the transmitted signals 
when reflecting off the human body. We define $r_L$ as the direct relation between the reflected $P_o$ and incident $P_i$ power values (in dB) on the target, $r_L = {P_o} - {P_i}$, where $P_i$ is the summation of the transmit power, the Tx gain and the path gain experienced by the plane wave at the target. $P_o$ represents the PDP at the target distance, with the exclusion of the Rx antenna gain and the corresponding path loss. Thus, the incident and reflected power values are no longer distance-dependent, as the two-way path loss effects have been eliminated. 

\subsection{Vital sign estimation}
\label{sec:vital_sign_est}
As discussed in Sec.~\ref{sec:pipelineoverview}, we separate the processing methods for the single- and two-person scenarios to analyse the capabilities of the mmWave communication multi-beam system for vital sign estimation. 
Once the time domain phase difference signal is converted to the frequency domain via the FFT, we use experimental results to show that in the single-person scenario, only one prominent peak related to a frequency tone is found, whereas, in the two-person scenario, the frequency domain signal shows prominent peaks related to more than one frequency tone. 
Hence, the DWT is employed in the single-person scenario to improve the estimation performance. 
Since the DWT cannot be used in multi-person scenarios for vital signs estimation, an FFT-based combination with the $k$-means clustering algorithm method is implemented to obtain vital signs from two humans with different vital sign activity. Prior to this, a subcarrier selection strategy is investigated to further increase performance based on frequency diversity occurring in the OFDM mmWave signals.

\subsubsection{Subcarrier analysis \& selection}

\begin{figure}[t]
\begin{subfigure}{0.49\textwidth}
        \centering      \includegraphics[width=\textwidth]{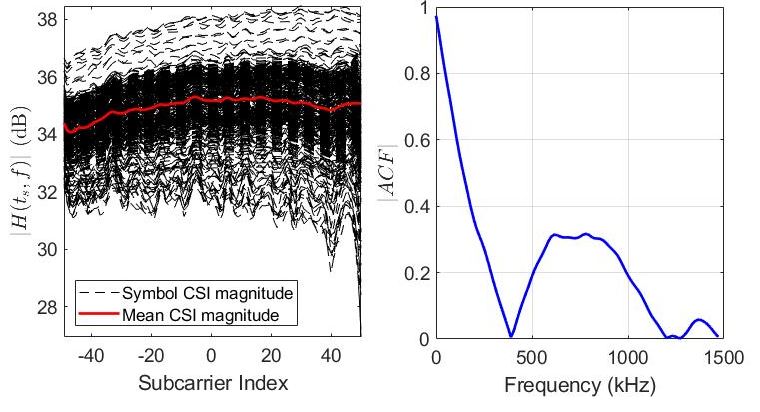}
             \caption{\centering{$\text{TS}_1$ (mono-static configuration)}}  
         \label{fig:varampa1}
     \end{subfigure}
    \vspace{0.2em}
    
    \begin{subfigure}{0.49\textwidth}
        \centering       \includegraphics[width=\textwidth]{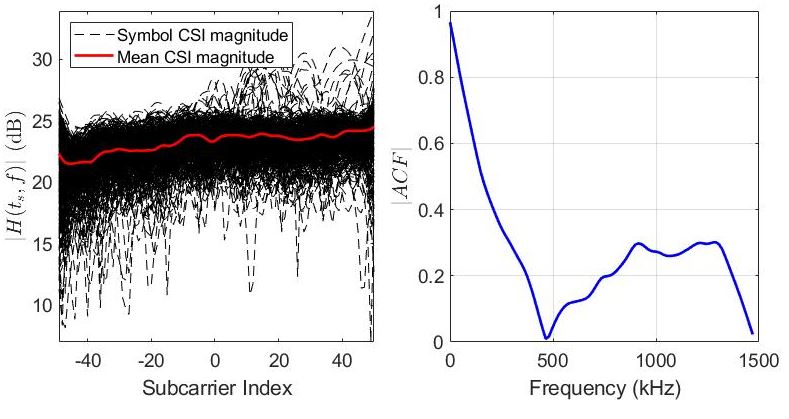}
             \caption{\centering{$\text{TS}_2$ (bistatic configuration)}}
         \label{fig:varampa2}
     \end{subfigure}
     \caption{\centering{Channel frequency response over transmitted symbols across subcarriers}}
    \label{fig:subvar}
\end{figure}

\begin{figure}[t]    
\centering
         \begin{subfigure}[b]{0.24\textwidth}
         \centering
         \includegraphics[width=\textwidth]{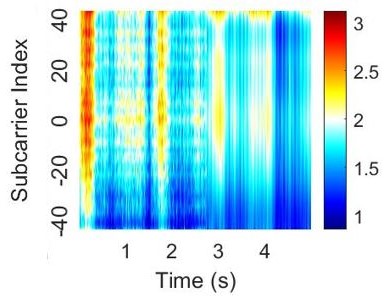}
         \caption{CSI magnitude}
         \label{fig:subvar_a}
     \end{subfigure}
     \hfill
         \begin{subfigure}[b]{0.24\textwidth}
         \centering
         \includegraphics[width=\textwidth]{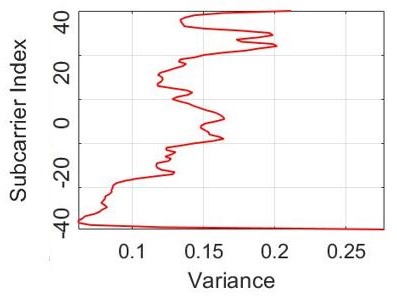}
         \caption{Time-domain variance}
         \label{fig:subvar_b}
     \end{subfigure}

         \caption{\centering{
         Channel autocorrelation function in $\text{TS}_1$}}
    \label{fig:subvar}
\end{figure}

During measurement data processing, it has been observed that the magnitude and phase of the CSI samples present variations across different subcarriers, meaning that subcarriers present different sensitivity to chest displacements and body movement because they have slightly different central frequencies, and the subcarrier separation is much smaller compared to the breathing/heartbeat cycle.
Fig.~\ref{fig:subvar} plots the frequency response of the channel across subcarriers for all transmitted symbols during the 5~s capture time, alongside the averaged channel response of all transmitted symbols, for the pair of Tx beam 9 and Rx beam 9. 
The plots present an overall trend of a frequency flat channel response, mainly due to the transmitted signal that reflects off the human chest and arrives back at the Rx without encountering other scatterers in its path. 
Nevertheless, for some specific symbols, the channel presents a frequency-selective behavior, with a deep fade up to 10 dB attenuation below and a peak up to 10 dB gain above the mean value. It is noticed that the mentioned peaks mainly occur in the higher frequency subcarriers. 

The autocorrelation function (ACF) of the channel is plotted in the right figures of Fig.~\ref{fig:subvar} (a) and (b). In both cases, we see that the 50\% coherence bandwidth is 135 kHz and 150 kHz for two plotted scenarios, respectively, which is much larger than the 15 kHz subcarrier width. The 90\% coherence bandwidth is at around 15 kHz for both configurations, hence further validating the flat fading behavior of the subcarrier channel in both tested configurations. 
The CSI magnitude pattern is plotted alongside the time variance per subcarrier in Fig.~\ref{fig:subvar} (c). It is well observed that subcarriers with larger variances are more sensitive to breathing cycles than others. Subcarriers at a higher frequency and those in the vicinity of the central frequency, $n_f = \left[-10...10,30...40 \right]$, present larger variance compared to those in the lower end of the bandwidth. Moreover, in the surface plot in Fig.~\ref{fig:subvar} (c) it is also noticeable that 4 breathing cycles occur during the 5~s capture duration, corresponding to the rise and fall of the chest.

\begin{figure}[t]
    \centering
    \includegraphics[width=0.48\textwidth]{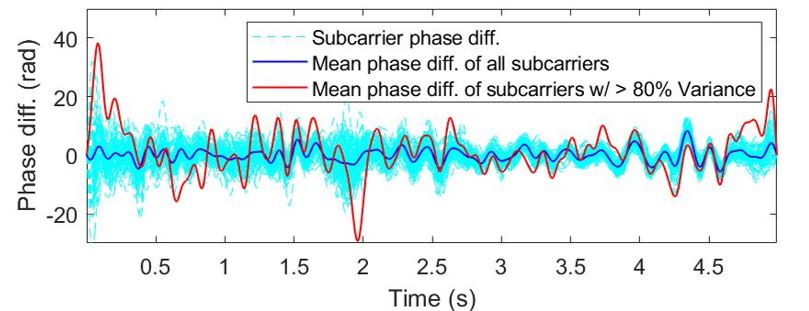}
    \caption{\centering{Effect of subcarrier selection on phase difference data for pair Tx beam 9 -- Rx beam 8}}
    \label{fig:pdiffVar}
\end{figure}
The above-observed subcarrier frequency behavior of the channel can be explained using physics mechanisms. In contrast to the specular reflection where the incident signal is reflected in a single outgoing direction, diffuse reflection occurs when a flat wave is scattered into multiple (random) directions due to interaction with a rough surface~\cite{gan}; the condition of the rough surface is relative to wavelength.
In~\cite{roughness}, the roughness of the human body is measured for slim, muscular and obese body types. It is found that the body surface roughness at the chest oscillates from 15.2~mm to 12.8~mm. The scattering of 26~GHz signals at the human chest surface can be considered as diffuse scattering in which the incident signal is scattered in different random directions. The diffuse scattering with random phase generates destructive and constructive interference at the Rx, which translates into the fades and peaks that appear in the channel frequency response for different transmitted symbols. It is therefore necessary to filter out subcarriers with low sensitivity and keep those with larger variance values to make sure the estimation is performed on subcarriers carrying useful information on the chest displacement activity.
To overcome this, a simple threshold detection method is used, where the threshold value is determined based on the measurement campaign. 
Fig.~\ref{fig:pdiffVar} shows an example of the threshold-based subcarrier selection.

\subsubsection{Single-person vital sign estimation}
Here we utilize the DWT to recursively decompose the phase difference data into an approximation coefficient vector with a low-pass filter and a detail coefficient vector with a high-pass filter. 
The approximation coefficient vector represents the low-frequency information of the input signal, while the detail coefficient vector describes the high-frequency detailed information \cite{phasebeat}. 
In the DWT theory, after $L$ steps, the DWT obtains the approximation coefficient $a^L$ and the sequence of detailed coefficients $\beta^1...\beta^L$ \cite{dwt}. 
Moreover, the sampling rate is halved after each decomposition step. Once the last level approximation and detailed coefficients are obtained, a peak search followed by a peak-to-peak time interval is performed in the obtained time domain decomposed signal for all selected subcarriers. Let $P_S$ be the number of selected subcarriers, then a set of peak-to-peak intervals from all $P_S$ subcarriers can be expressed as $L = \left[ l_1,...,l_{P_S} \right]$, where $l_i$ is the mean value of the vector containing the $N$ peak-to-peak intervals obtained from the $i^{\text{th}}$ subcarrier, where $i\in 1...P_s$. Even if all selected subcarriers are above the measurement-based variance threshold, some of them may be more prominent than others. Therefore, we compute a weighted mean of $L$ by taking the variance of each subcarrier into account. The final estimation value is obtained as
\begin{equation}
    E = \sum_{i=1}^{P_S}\frac{v_i\cdot l_i}{\sum_{i=1}^{P_S}v_i},
\end{equation}
where $v_i$ is the variance of the $i^{\text{th}}$ subcarrier. 

As shown in Fig.~\ref{fig:pipeline}, after selecting the prominent subcarriers carrying larger phase variations across the time samples, the processed phase difference data is input to a Finite Impulse Response (FIR) bandpass filter with a pass band of the typical frequency range of breathing and heartbeat rates. We select a frequency range of 0.08 to 1 Hz (4.5 to 60 bpm) for the breathing rate (BR), and a frequency range of 1 to 2 Hz (60 to 120 bpm) for the heartbeat rate (HR), covering both normal and abnormal breathing/heartbeat rates~\cite{lung}. Afterwards, an FFT is performed in the time domain phase difference data. Let us denote $\omega_b$ as the selected frequency range for BR, $\omega_h$ as the corresponding frequency range for HR, and $\phi(t)$ as the time domain phase difference data. The BR and HR estimates are then computed as:
\begin{equation}
    \hat{f}_{b} = \arg \max_{f\in \omega_b} \left|\text{FFT}(\phi(t)) \right|,
\end{equation}
\begin{equation}
    \hat{f}_{h} = \arg \max_{f\in \omega_h} \left|\text{FFT}(\phi(t)) \right|.
\end{equation}

\subsubsection{Multi-person vital sign estimation}
For the multi-person scenario, an FFT-based $k$-means clustering algorithm in the frequency domain for selected subcarriers is proposed to obtain vital sign estimates of two persons with different breath- and heartbeat rates. 
The number of cluster targets in the measurement scenario is estimated as the number of peaks in the frequency domain signal that is above a certain power threshold. This value serves as input to the $k$-means algorithm. The aforementioned threshold will be later discussed and is based on experimental and numerical analysis. For a given selected subcarrier $n_f$, a peak search is performed to obtain peaks of the signal above a threshold. $k$-means clustering is then performed on the selected peaks. The centroid obtained from the clustering is used as the BR and/or HR estimate at that subcarrier. Given the set of centroids $C = [c_1,...,c_{P_S}]$ from all selected subcarriers, the final estimated rate is obtained by computing the weighted mean of $C$ using subcarrier variance as the weighting coefficients.

\section{Performance Evaluation}
\label{sec:numanalysis}

In this section, we evaluate the performance of using the 26~GHz multi-beam communication testbed and the proposed pipeline for contact-free single and multi-person vital sign estimation. The measurement setup and scenarios have been introduced in Sec.~\ref{sec:system_experiment}.

\subsection{Channel amplitude \& reflection loss analysis}

\begin{figure}[t]
     \centering
     \begin{subfigure}[b]{0.24\textwidth}
         \centering
         \includegraphics[width=\textwidth]{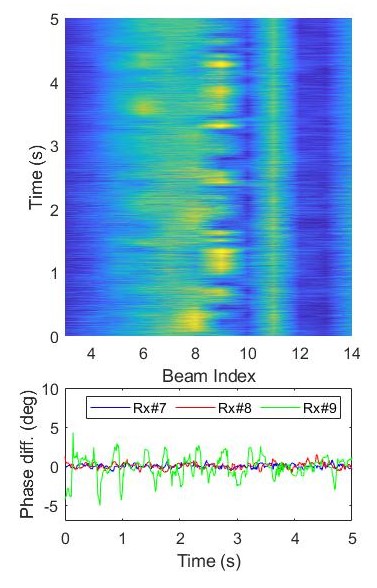}
         \caption{Tx beam 8}
         \label{fig:csiamp1}
     \end{subfigure}
     \hfill
     \begin{subfigure}[b]{0.235\textwidth}
         \centering
         \includegraphics[width=\textwidth]{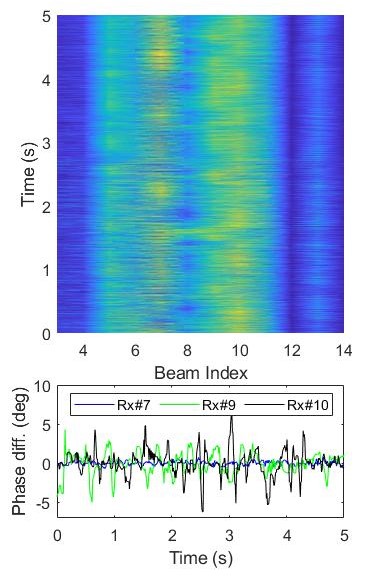}
         \caption{Tx beam 9}
         \label{fig:csiamp2}
     \end{subfigure}
     \caption{Calibrated CSI amplitude data across Rx beams.}
     \label{fig:csiamp}
\end{figure}

We show the amplitude of the CSI samples over time and across Rx beams for the two Tx beams for $\text{TS}_1$ in Fig.~\ref{fig:csiamp}. The top figures 
correspond to the amplitude variations over time, while the bottom figures show the extracted and processed phase difference data of the beams with larger CSI amplitude. First, both figures present the channel amplitude variations over time, most likely due to the multipath components arriving from different directions and causing destructive interference to the received signal at a specific temporal sample. Rx beams 9 and 10 possess larger amplitudes when transmitting with Tx beams 9, while Rx beams 8 and 9 possess larger amplitudes when transmitting with Tx beams 8. In the bottom phase difference plots, it can be observed that the aforementioned Rx beams possess larger phase variations related to vital sign activity than others with lower amplitude levels. This observation proves that beam selection is needed for a better estimation of vital sign. In Sec.~\ref{sec:vs1p} the estimation performance over Rx beams will be further addressed.

\begin{figure}[t]
    \centering
    \includegraphics[width = 0.48\textwidth]{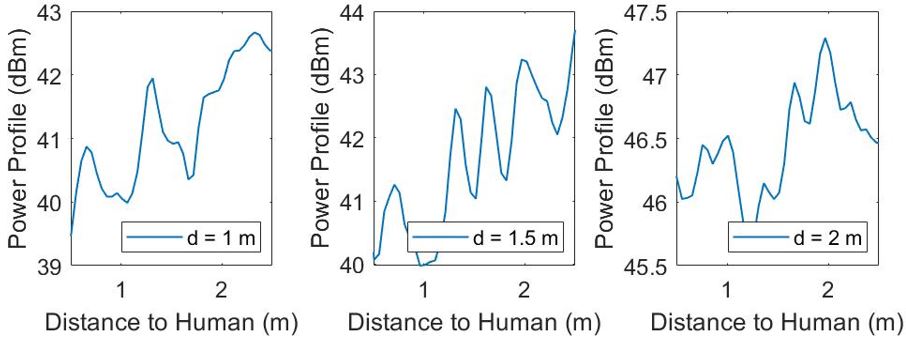}
    \caption{Mean PDP for different Tx-target distances}
    \label{fig:pdp}
\end{figure}

The mean PDP of the two transmitting and reception beam pairs for $\text{TS}_1$ scenario with the distance between Tx and human target are 1~m, 1.5~m and 2~m are shown in Fig.~\ref{fig:pdp}. The two central beams with a 7\textdegree~beamwidth: the pair of Tx beam 8 - Rx beam 8 and the pair of Tx beam 9 - Rx beam 8, were used in transmission point towards the target's chest. In Fig.~\ref{fig:pdp}, the prominent peaks are observable at the distances where the human target was positioned. When the target was seated at 1 m, 1.5 m and 2 m away from the Tx/Rx module, a prominent peak is observed at the equivalent delay distance of 1.3 m, 
1.7 m and 2.1 m respectively. 

\begin{figure}[t]
    \centering
    \includegraphics[width = 0.25\textwidth]{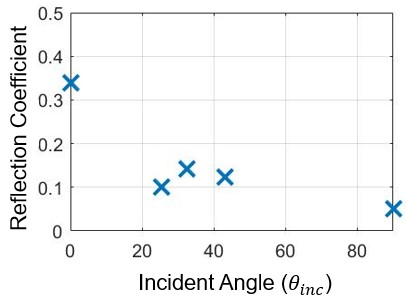}
    \caption{Impact of incident angle on
reflected power}
    \label{fig:rloss}
\end{figure}

For both $\text{TS}_1$ and $\text{TS}_2$ scenarios, we compute the reflection loss due to human backscattering for the selected observing Rx beams to ensure reflections are coming from the human target. Fig.~\ref{fig:rloss} shows the resulting reflection coefficient values. 
Around 35\% of the power is reflected off the human chest at a normal incident angle. The reflection coefficient decreases as the incident angle increases. When the incident angle increases, fewer reflections are captured at the Rx beams. These results are in accordance with the values obtained in \textit{Wu et al.}~\cite{rloss}, which is based on simulated theoretical models. 

\subsection{Single-person Vital Sign Estimation}

We have observed in Fig.~\ref{fig:calproc} how the raw CSI phase difference data presents very fast variations. After phase calibration to remove the hardware non-linearity, we implement a Hampel filter to remove the DC trend and perform down-sampling to acquire a more manageable data size. In this work, a down-sampling factor of 20 is implemented to bring the original 2000 Hz sampling frequency and 10000 time samples down to 100 Hz sampling frequency and 500 time samples, respectively. 

\begin{figure}[tb]
     \centering
     \begin{subfigure}[b]{0.48\textwidth}
         \centering
         \includegraphics[width=\textwidth]{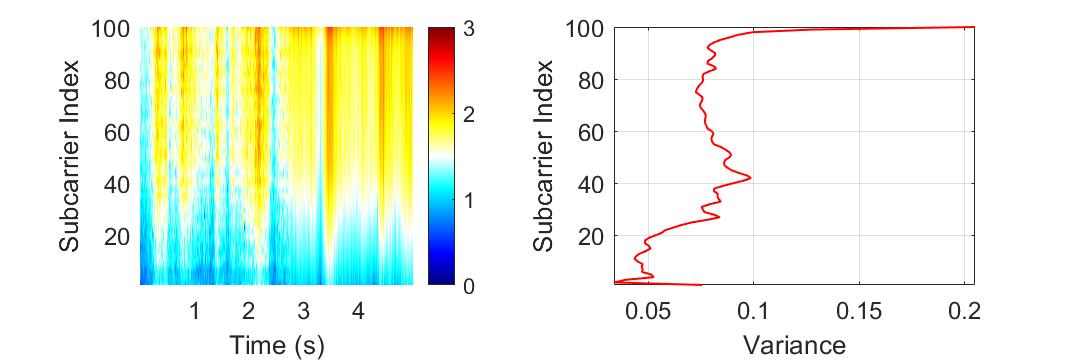}
         \caption{TX beam 8 -- Rx beam 8}
         \label{fig:var1}
     \end{subfigure}
     \hspace{0.1em}
     
     \begin{subfigure}[b]{0.48\textwidth}
         \centering
         \includegraphics[width=\textwidth]{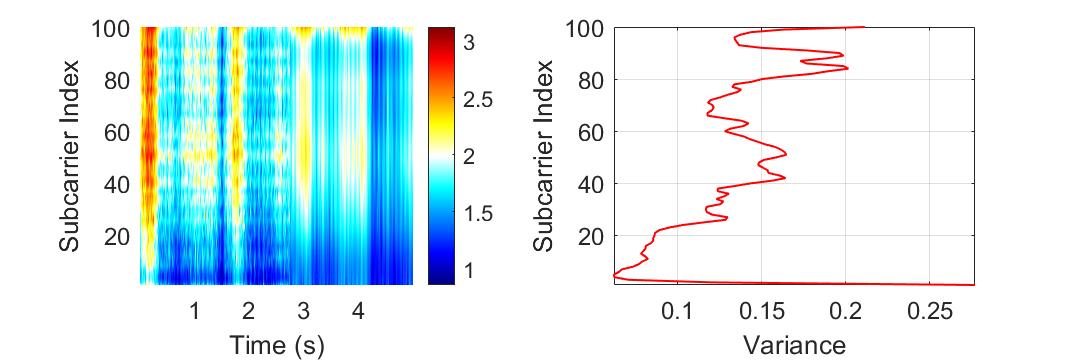}
         \caption{TX beam 9 -- Rx beam 8}
         \label{fig:var2}
     \end{subfigure}
     \caption{\centering{Variance across time and subcarrier index for central beam pair links in $\text{TS}_2$}}
     \label{fig:varvstime}
\end{figure}

Subcarrier selection is an important step in vital sign estimation. In Fig.~\ref{fig:pdiffVar}, individual subcarrier phase differences are plotted and compared to the mean phase difference across all subcarriers, as well as the mean phase difference of subcarriers with $>$ 80\% of the maximum variance across the time domain. 
In Fig.~\ref{fig:varvstime}, the left column shows the surface plot of CSI amplitude samples versus time and the right column depicts time domain variance versus subcarrier index for different beam pairs. 

\begin{figure}[tb]
    \centering
    \includegraphics[width= 0.48\textwidth]{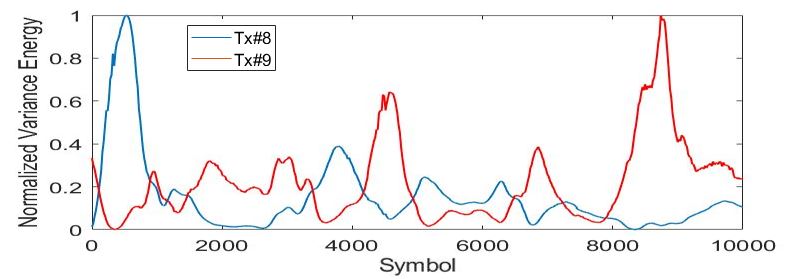}
    \caption{Normalized variance energy at Rx beam 8}
    \label{fig:varenergy}
\end{figure}

The impact of chest displacement due to chest activity is particularly noticeable in Fig.~\ref{fig:varenergy}. The normalized variance energy per beam is defined as
\begin{equation}
    NVE = \frac{1}{N_fN_s}\sum_{n_f = 1}^{N_f} \sum_{t_s = 1}^{N_s}\left | v_{n_f}(t_s) \right |,
    \label{eq:energy}
\end{equation}
where $v_{n_f}(t_s)$ is the variance value at subcarrier $n_f$ and symbol $t_s$, obtained from the variance matrix $\bm{V} = \left[ V_1,...,V_{N_f} \right ]$ with $V_{n_f} = \left[ v_1(1),...,v_{N_f}(N_s) \right ]^{T}$. Variance values per subcarrier are spliced in time sample windows at which the cumulative variance energy is obtained. The results show three peaks where the variance energy is concentrated, corresponding to the three breathing cycles performed by the person.

\begin{figure}[t]
     \centering
     \begin{subfigure}{0.23\textwidth}
         \centering
         \includegraphics[width=\textwidth]{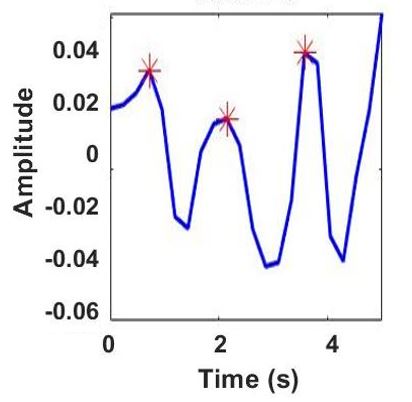}
         \caption{Breathing pattern}
         \label{fig:dwtb}
     \end{subfigure}
     \hspace{0.1em}
     \begin{subfigure}{0.218\textwidth}
         \centering
         \includegraphics[width=\textwidth]{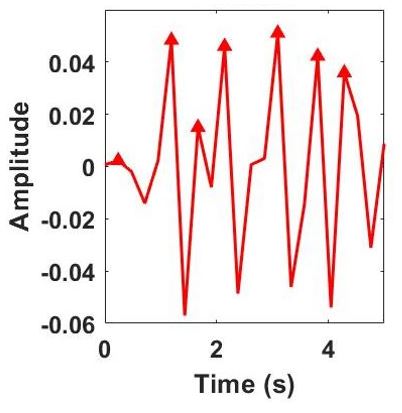}
         \caption{Heartbeat pattern}
         \label{fig:dwth}
     \end{subfigure}
     \caption{\centering{DWT signals for the pair Tx beam 9 -- Rx beam 9}}
     \label{fig:dwt}
\end{figure}

\begin{figure}[t]
    \centering
    \includegraphics[width=0.48\textwidth]{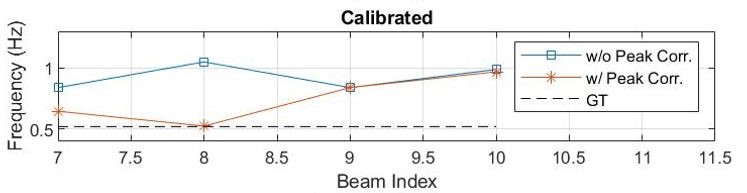}
    \caption{\centering{False peak detection and removal example using subcarrier selection}}
    \label{fig:estcorr}
\end{figure}

Once the data is processed, we employ a DWT with $L$ = 4 steps and Daubechies wavelet filters of level 4. This value is chosen to scale down the sampling frequencies to our desired range; with each step, we half the sampling frequencies.
Fig.~\ref{fig:dwt} depicts the extracted DWT breathing and heartbeat signals for the pair of Rx beam 9 and Tx beam 8. The time positions of the peaks are subtracted from their respective following peak, to obtain the period $P$. The rate is then computed in Hz as 1/$P$ or 60/$P$ in bpm. The ground-truth rates are 0.56 Hz for breath and 1.37 Hz for heartbeat activities when using the on-body sensors as references. The prior knowledge of typical breathing and heartbeat ranges of humans is leveraged to remove falsely detected peaks on the DWT. Such peaks that present inter-peak interval times outside the range of typical values are deleted. Estimation results shown in Fig.~\ref{fig:estcorr} indicate the importance of the false peak removal method for the breathing cycle. 
The estimation using the received signal of Rx beam 8 achieves the best result as expected, as the beam is directly pointing to the human chest.

\begin{figure}[t]
    \centering
     \begin{subfigure}[b]{0.23\textwidth}
         \centering
         \includegraphics[width=\textwidth]{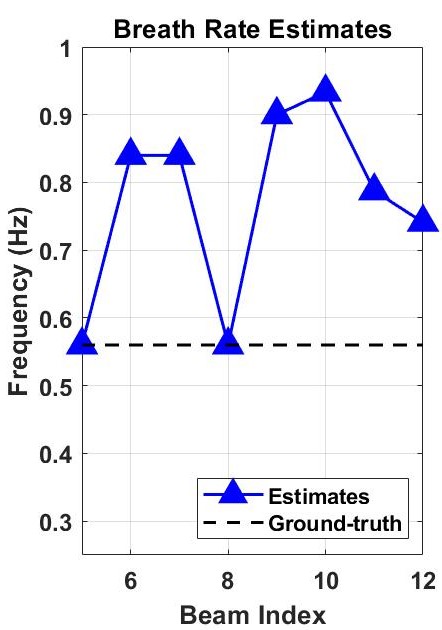}
         \caption{Breath rate}
         \label{fig:est1_a}
     \end{subfigure}
     \hspace{0.1em}
     \begin{subfigure}[b]{0.23\textwidth}
         \centering
         \includegraphics[width=\textwidth]{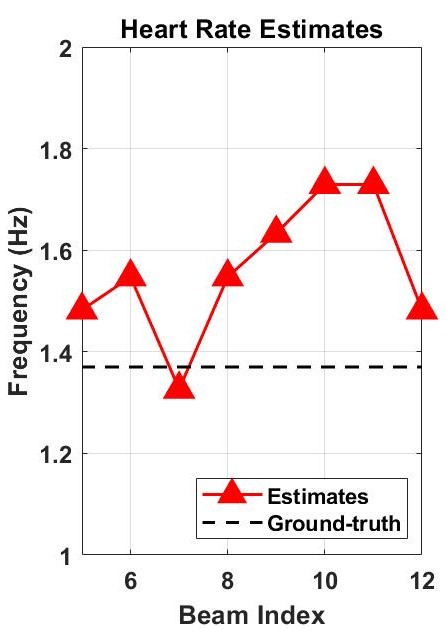}
         \caption{Heart rate}
         \label{fig:est1_b}
     \end{subfigure}
    \caption{\centering{
    DWT-based estimation results for $\text{TS}_1$ with the human placed 3 m from the Tx, using Tx beam 9 and Rx beam 5 to 12}}
    \label{fig:est1}
\end{figure}

An example of DWT-based estimation results for Rx beams 5-12 is shown in Fig.~\ref{fig:est1}. Rx beams 5 and 8 achieve perfect breath rate estimation compared to the ground truth. This is expected for Rx beam 8 as the beam is directly pointing to the human chest. For the case of Rx beam 5, results show that signals reflecting in the human chest can also be picked up from different azimuth angles. For heartbeat rate estimates, Rx beam 7 achieves 1.2 bpm absolute error. Overall, from Fig.~\ref{fig:est1}, it is shown that the estimation is very poor for those Rx beams that can not pick up contributions from signals reflected off the human chest, but highly accurate from the receive beams picking up reflections from the human chest due to their directivity towards human directions or multi-scattering. In real-life applications, beam selection is essential to choose the information picked up from Rx beams whose FoV covers the human's chest. 

Moreover, there is a clear overestimation for all beams, thus, affecting the breathing and heart rate estimates in a similar way. An overestimation of the frequency rate translates into an underestimation of the breathing and heartbeat periods that are extracted from the DWT decomposed low- and high-frequency signals in the band of interest. It means that the obtained mean inter-peak interval is actually smaller than the ground truth. The DWT decomposition on beams not carrying information on the vital sign produces false peaks with small amplitudes; these false peaks do not correspond to vital sign activity but are still considered after the correction mechanism because they are still inside the bounds of the vital sign periods used in the correction step. Another reason is that in the 5s of capture duration, there are around 3 and 7 full breathing and heartbeat cycles, respectively, meaning that the average inter-peak interval has a very small number of samples, amplifying the importance of the DWT method to precisely extract the correct frequency tones.

\begin{figure}[t]
     \centering
     \begin{subfigure}[b]{0.24\textwidth}
         \centering
         \includegraphics[width=\textwidth]{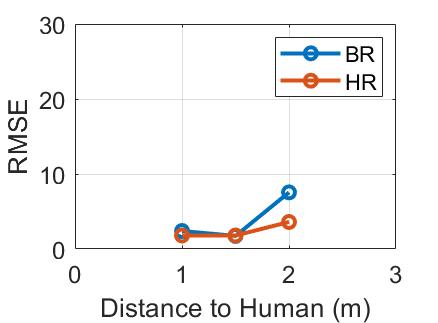}
         \caption{Various Tx--human distances}
         \label{fig:estvsd}
     \end{subfigure}
     \hfill
     \begin{subfigure}[b]{0.24\textwidth}
         \centering
         \includegraphics[width=\textwidth]{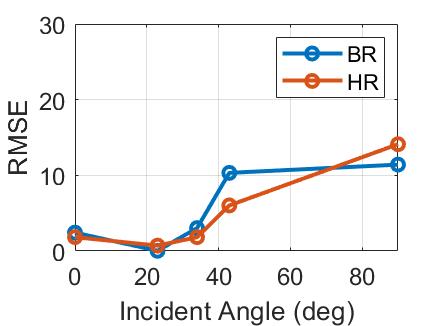}
         \caption{Various incident angles}
         \label{fig:estvsincang}
     \end{subfigure}
     \caption{\centering{Estimation RMSE values for the pair Tx beam 9 -- Rx beam 8}}
     \label{fig:rmse}
\end{figure}

The RMSE values of vital sign estimation with respect to distance and incident angles are plotted in Fig.~\ref{fig:rmse} for $\text{TS}_1$ and $\text{TS}_2$. 
These values are obtained by the DWT method for each Rx beam and selecting the beam with a smaller RMSE value. Hence, the results in Fig.~\ref{fig:rmse} represent the best-performing Rx beam. 
Although not having a very thorough data set, we observe a general trend that the estimation error increases as both distance and incident angle increase. A reason for this is that at larger distances, the reflected signals are more attenuated, while at large incident angles, the reflected signals affected by the human chest may be not picked up by beams with larger gains. The RMSE values for breathing rate and heartbeat rate are kept under 5 bpm for all the measured distances and for the incident angles below 40 degrees.

\label{sec:vs1p}

 \par

\par
\par

\begin{figure}[t]
     \centering
     \begin{subfigure}[b]{0.24\textwidth}
         \centering
         \includegraphics[width=\textwidth]{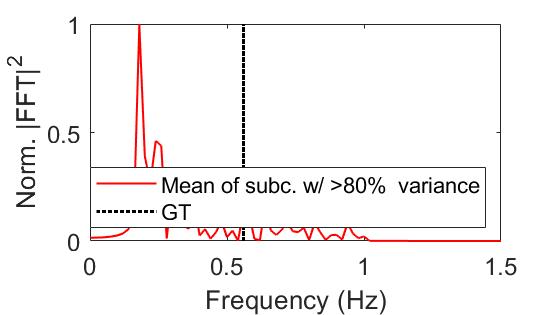}
         \caption{FFT-based method}
         \label{fig:spfft}
     \end{subfigure}
     \hfill
     \begin{subfigure}[b]{0.24\textwidth}
         \centering
         \includegraphics[width=0.85\textwidth]{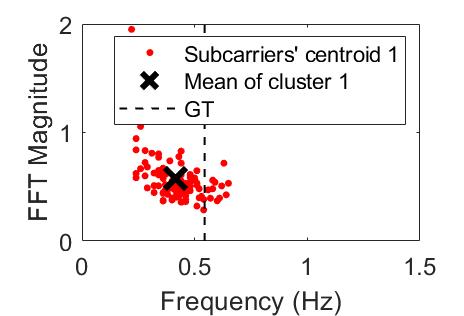}
         \caption{FFT with k-means clustering}
         \label{fig:spfft1}
     \end{subfigure}
     \caption{\centering{
     FFT-based single person estimate using the pair Tx beam 9 -- Rx beam 8}}
     \label{fig:spfftt}
\end{figure}

The FFT-based method depicted in Sec.~\ref{sec:method} is used to compare with the proposed DWT-based method. Fig.~\ref{fig:spfftt} (a) shows the frequency domain mean signal power of subcarriers with $>$80\% of the maximum variance value of the Tx beam 9 - Rx beam 8 pair link for $\text{TS}_1$ with the target positioned at 1~m from the Tx. Typical FFT-based methods for vital sign estimation found in the literature (see Sec.\ref{sec:stateoftheart}) rely on selecting the maximum peak; in Fig.~\ref{fig:spfftt} (a), we obtain an absolute error of more than 20 bpm by comparing the estimated rate (shown as peaks in the figure) to the ground-truth rate. 
In Fig.~\ref{fig:spfftt} (b) the selected subcarriers' centroids are depicted in red dots. The mean weighted average of these points is shown as the black cross, taking the variance of each subcarrier as the weighting factor. We observe an absolute error of 0.1435 Hz (8.61 bpm) compared to the ground truth, which is $>$7 bpm worse than the one obtained with the proposed DWT method for the same Tx-Rx pair link.

\subsection{Multi-person vital sign estimation}
\begin{figure}[t]
    \centering
    \begin{subfigure}[b]{0.24\textwidth}
         \centering
         \includegraphics[width=\textwidth]{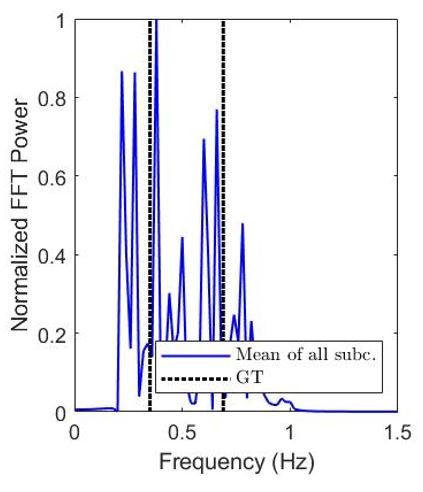}
         \caption{Tx beam 9 -- Rx beam 6}
         \label{fig:vs2p1}
     \end{subfigure}
     \hfill
     \begin{subfigure}[b]{0.24\textwidth}
         \centering
         \includegraphics[width=\textwidth]{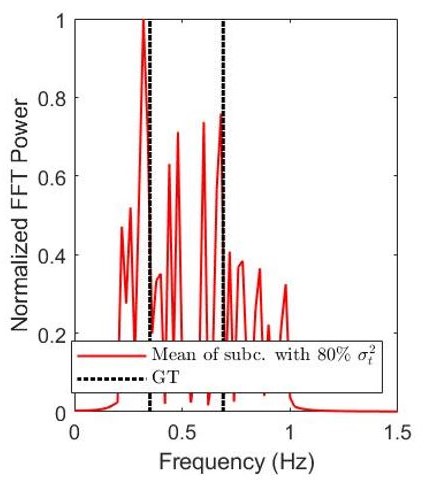}
         \caption{Tx beam 9 -- Rx beam 12}
         \label{fig:vs2p2}
     \end{subfigure}

    \caption{\centering{FFT analysis of the two-person scenario
    for the pair Tx beam 9 -- Rx beam 12}}
    \label{fig:fft2p}
\end{figure}

The vital sign of two-person are estimated through $\text{TS}_3$ scenario, in which 
the two measured targets breathe at different rates. The ground-truth breath rate values for target $H_1$ and $H_2$ are 0.35 Hz (21 bpm) and 0.69 Hz (41.1 bpm), respectively. Firstly, let us compare the normalized frequency power illustrated in Fig.~\ref{fig:spfftt} (a) for single-person and Fig.~\ref{fig:fft2p} for two-person. Compared to the single-person scenario, two prominent peaks are seen in the two-person scenario. We exploit this finding to determine the number of target inputs to the $k$-means algorithm for estimating the rates. 

\begin{figure}[t]
     \centering
     \begin{subfigure}[b]{0.24\textwidth}
         \centering
         \includegraphics[width=\textwidth]{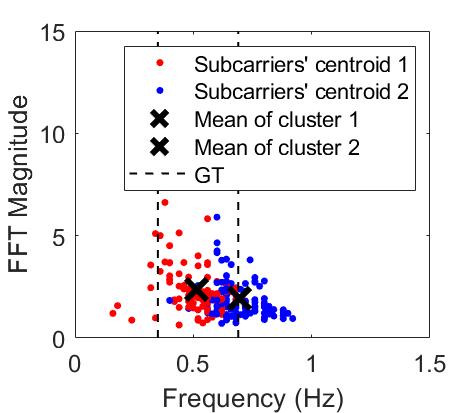}
         \caption{Tx beam 9 -- Rx beam 6}
         \label{fig:vs2p1}
     \end{subfigure}
     \hfill
     \begin{subfigure}[b]{0.24\textwidth}
         \centering
         \includegraphics[width=\textwidth]{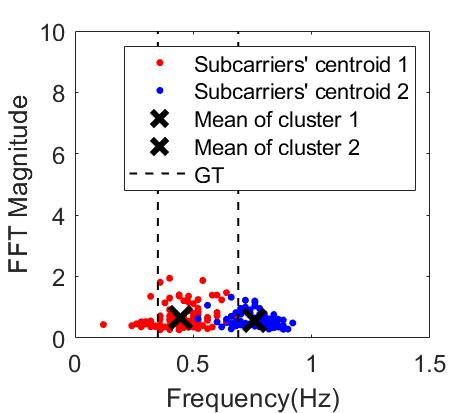}
         \caption{Tx beam 9 -- Rx beam 12}
         \label{fig:vs2p2}
     \end{subfigure}
          \centering
     \begin{subfigure}[b]{0.24\textwidth}
         \centering
         \includegraphics[width=\textwidth]{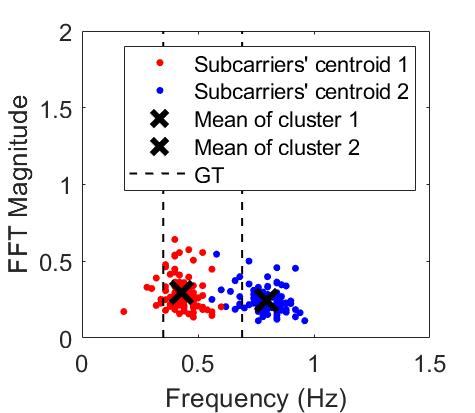}
         \caption{Tx beam 8 -- Rx beam 6}
         \label{fig:vs2p3}
     \end{subfigure}
     \hfill
     \begin{subfigure}[b]{0.24\textwidth}
         \centering
         \includegraphics[width=\textwidth]{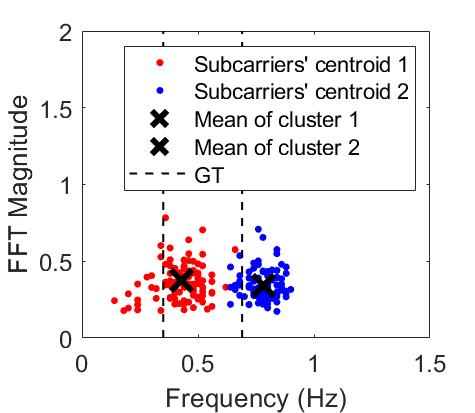}
         \caption{Tx beam 8 -- Rx beam 12}
         \label{fig:vs2p4}
     \end{subfigure}
          \caption{\centering{Breath rate estimates for the two-person scenario in $\text{TS}_3$}}
     \label{fig:vs2}
\end{figure}

The breathing rate estimation results are depicted in Fig.~\ref{fig:vs2}. Based on the geometry of $\text{TS}_3$ and the Butler matrix pattern, some insights can be drawn. 
Observing Fig.~\ref{fig:vs2}~(a) and Fig.~\ref{fig:vs2}~(b), in which Tx beam 9 is used, it is observable that estimation on target $H_1$ is poorer than the one for target $H_2$. It is because the beam used in transmission is mainly pointing to target $H_2$, hence the backscattering will affect the frequency to a larger extent. Nevertheless, the Tx beam is also capable of reaching target $H_1$, 
affecting the reflected signals with its frequency. In this case, Rx beam 6 and Rx beam 12 have 0 and 1.5 bpm error values, respectively. 

When transmitting with Tx beam 8, the error is kept under 2 bpm in all cases for both targets, as shown in Fig.~\ref{fig:vs2}~(c) and Fig.~\ref{fig:vs2}~(d). The presented results for the two-person non-frontal view scenario reveal that individual beams have the capability of recovering information from the backscattered signals in a bistatic indoor setting. The estimation performance relates directly to the layout of the scenario and the orientation of the Tx and Rx beams relative to the targets, which can also be leveraged to obtain vital sign estimation on targets breathing at similar rates but separated/covered by different beams. It is beyond the scope of this paper but will be investigated in future work. {A performance comparison between the proposed pipeline using the proposed multi-beam communication system in context of JCAS and the reviewed state-of-the-art performances/systems in Section \ref{sec:stateoftheart} can be found in Table \ref{tab:state_art}.}

\begin{table}[t]
\centering
\captionsetup{justification=centering, labelsep=newline}
\caption{{Benchmarking the performance of the proposed pipeline using the proposed multi-beam communication system in context of JCAS to the performances in literature review}}
\label{tab:state_art}
\resizebox{0.48\textwidth}{!}{%
\begin{tabular}{ll}
\hline
\textbf{Method} & \textbf{Error (BR)}\\ \hline
\textbf{WiFi} & 1 bpm up to 11 m \cite{wifi1}-\cite{breathtrack} \\ 
\textbf{UHF Tags}            & 1 bpm (8$\times$8 Tags) with 3 persons \cite{holo}      \\
\textbf{UWB}                 & 99\% accuracy with 3 persons up to 7 m \cite{mtrack}         \\
\textbf{mMWave Radar}       & 3 bpm up to 2 m (8 GHz) and 5 m (77 GHz) \cite{radar8, radar77}\\
\textbf{This work (monostatic cfg.)}             & 8 \textless BR \textless 12 bpm with 1 person up to 2m \\
\textbf{This work (bistatic cfg.)}             & \textless 3 bpm with 2 person up to 3 m \\ \hline
\end{tabular}}
\end{table}

{
Note that the proposed pipeline would still work for the two-person scenario when the two persons are on the same side of the direct path between BS and UE, or even for more-than-two-person scenario, if the transmit and receive beam pairs via the two persons are not overlapping. In other words, if there is one transmit beam impinging only on the first person and then received by one receive beam, and there is another transmit beam impinging only on the second person and then received by another receive beam, we could still utilize the same pipeline. This condition is fulfilled depending on the beam separation, the persons' location, body size and orientation. }


\section{Conclusion}
\label{sec:conclusions}
In this paper, the capability of a 26 GHz multi-beam OFDM communication testbed for vital signs estimation has been studied based on a measurement campaign on single and multi-person scenarios. A post-processing pipeline for vital sign estimation is introduced, implemented and evaluated. The pipeline includes the frequency and time domain calibrations of the captured CSI samples to remove frequency-phase nonlinearities introduced by the up/down conversion RF chain as well as the high-frequency denoising. We propose to use distinct approaches for estimating the breathing and heart rate for the single- and multi-person scenarios. A DWT is employed in single-person scenarios to obtain better estimates compared to FFT-based methods. For multi-person scenarios, an FFT and $k$-means clustering algorithm is employed to obtain vital sign information from different targets. The influence of frequency/subcarrier diversity and fading is also studied during the design of the pipeline. This work is mission-critical for 6G joint communication and sensing, providing proof-of-concept of using a multi-beam communication system for sensing vital signs.

From our numerical analysis, it is found that the reflection is larger at the normal (to chest surface) incident angle with 35\% of the incident power reflected. 
The proposed methodology in a monostatic radar-like BS/AP sensing scenario reveals that DWT can offer estimation performance below 2 bpm absolute error for individual Tx and Rx beam pair links while using FFT-based methods the error can reach up to 8 bpm. 
In $\text{TS}_2$ with separated Tx/Rx, the estimation error increases with the incident angle. 
In the two-person scenario, it is shown that individual beams are able to pick up reflections coming from different targets while offering a 3 bpm absolute error. 
It is a promising indication that a multi-beam base station or access point could sense the vital signs of multiple persons while communicating with users, given that the targets and users are covered by different beams.

Future lines of work should include a more extensive measurement campaign 
to provide an insight into the influence of distance and incident angle in both reflection/backscattering and vital signs estimation. 
Such campaign should make it possible to better evaluate the proposed pipeline in real-world indoor scenarios with more than two targets, and simultaneously evaluate the capability of the system for joint active communication and sensing activities. Based on the spatial diversity of the multi-beam system, mapping targets in space based on their vital signs could also be an interesting application in a future line of work.


\bibliographystyle{IEEEtran}
\bibliography{IEEEfull,Bibliography}

\begin{thebibliography}{10}
\providecommand{\url}[1]{#1}
\csname url@samestyle\endcsname
\providecommand{\newblock}{\relax}
\providecommand{\bibinfo}[2]{#2}
\providecommand{\BIBentrySTDinterwordspacing}{\spaceskip=0pt\relax}
\providecommand{\BIBentryALTinterwordstretchfactor}{4}
\providecommand{\BIBentryALTinterwordspacing}{\spaceskip=\fontdimen2\font plus
\BIBentryALTinterwordstretchfactor\fontdimen3\font minus \fontdimen4\font\relax}
\providecommand{\BIBforeignlanguage}[2]{{%
\expandafter\ifx\csname l@#1\endcsname\relax
\typeout{** WARNING: IEEEtran.bst: No hyphenation pattern has been}%
\typeout{** loaded for the language `#1'. Using the pattern for}%
\typeout{** the default language instead.}%
\else
\language=\csname l@#1\endcsname
\fi
#2}}
\providecommand{\BIBdecl}{\relax}
\BIBdecl

\bibitem{6G_WhitePaper_Yang}
\BIBentryALTinterwordspacing
A.~Bourdoux, A.~N. Barreto, B.~van Liempd, C.~de~Lima, D.~Dardari, D.~Belot, E.-S. Lohan, G.~Seco-Granados, H.~Sarieddeen, H.~Wymeersch, J.~Suutala, J.~Saloranta, M.~Guillaud, M.~Isomursu, M.~Valkama, M.~R.~K. Aziz, R.~Berkvens, T.~Sanguanpuak, T.~Svensson, and Y.~Miao, ``{6G White Paper on Localization and Sensing},'' 2020. [Online]. Available: \url{https://arxiv.org/abs/2006.01779}
\BIBentrySTDinterwordspacing

\bibitem{10005804}
C.~Li, S.~D. Bast, Y.~Miao, E.~Tanghe, S.~Pollin, and W.~Joseph, ``{Contact-Free Multi-Target Tracking Using Distributed Massive MIMO-OFDM Communication System: Prototype and Analysis},'' \emph{IEEE Internet of Things Journal}, pp. 1--1, 2023.

\bibitem{https://doi.org/10.48550/arxiv.2209.08847}
\BIBentryALTinterwordspacing
H.~Alidoustaghdam, A.~Kokkeler, and Y.~Miao, ``{Multibeam Sparse Tiled Planar Array for Joint Communication and Sensing},'' 2022. [Online]. Available: \url{https://arxiv.org/abs/2209.08847}
\BIBentrySTDinterwordspacing

\bibitem{9743500}
H.~Alidoustaghdam, Y.~Miao, and A.~Kokkeler, ``{Integrating TDD Communication and Radar Sensing in Co-Located Planar Array: A Genetic Algorithm Enabled Aperture Design},'' in \emph{2022 2nd IEEE International Symposium on Joint Communications and Sensing (JCS)}, 2022, pp. 1--6.

\bibitem{9924658}
C.~Li, S.~De~Bast, Y.~Miao, E.~Tanghe, S.~Pollin, and W.~Joseph, ``{Contact-Free Pedestrian Tracking Using Massive MIMO-OFDM Communication System},'' in \emph{2022 19th European Radar Conference (EuRAD)}, 2022, pp. 181--184.

\bibitem{10001163}
R.~Hersyandika, Y.~Miao, and S.~Pollin, ``{Guard Beam: Protecting mmWave Communication through In-Band Early Blockage Prediction},'' in \emph{GLOBECOM 2022 - 2022 IEEE Global Communications Conference}, 2022, pp. 4093--4098.

\bibitem{10000833}
B.~van Berlo, Y.~Miao, R.~Hersyandika, N.~Meratnia, T.~Ozcelebi, A.~Kokkeler, and S.~Pollin, ``{Intelligent Blockage Recognition using Cellular mmWave Beamforming Data: Feasibility Study},'' in \emph{GLOBECOM 2022 - 2022 IEEE Global Communications Conference}, 2022, pp. 4576--4582.

\bibitem{jcas1}
T.~Wild, V.~Braun, and H.~Viswanathan, ``{Joint Design of Communication and Sensing for Beyond 5G and 6G Systems},'' \emph{IEEE Access}, vol.~9, pp. 30\,845--30\,857, 2021.

\bibitem{jcas2}
S.~D. Liyanaarachchi, T.~Riihonen, C.~B. Barneto, and M.~Valkama, ``{Optimized Waveforms for 5G–6G Communication With Sensing: Theory, Simulations and Experiments},'' \emph{IEEE Transactions on Wireless Communications}, vol.~20, no.~12, pp. 8301--8315, 2021.

\bibitem{jcas4}
C.~B. Barneto, S.~D. Liyanaarachchi, T.~Riihonen, L.~Anttila, and M.~Valkama, ``{Multibeam Design for Joint Communication and Sensing in 5G New Radio Networks},'' in \emph{ICC 2020 - 2020 IEEE International Conference on Communications (ICC)}, 2020, pp. 1--6.

\bibitem{8805161}
C.~Baquero~Barneto, T.~Riihonen, M.~Turunen, L.~Anttila, M.~Fleischer, K.~Stadius, J.~Ryynänen, and M.~Valkama, ``{Full-Duplex OFDM Radar With LTE and 5G NR Waveforms: Challenges, Solutions, and Measurements},'' \emph{IEEE Transactions on Microwave Theory and Techniques}, vol.~67, no.~10, pp. 4042--4054, 2019.

\bibitem{scholkmann2019pulse}
F.~Scholkmann and U.~Wolf, ``{The pulse-respiration quotient: A powerful but untapped parameter for modern studies about human physiology and pathophysiology},'' \emph{Frontiers in physiology}, vol.~10, p. 371, 2019.

\bibitem{intro1}
M.~S. Mahmud, H.~Wang, A.~M. Esfar-E-Alam, and H.~Fang, ``{A Wireless Health Monitoring System Using Mobile Phone Accessories},'' \emph{IEEE Internet of Things Journal}, vol.~4, no.~6, pp. 2009--2018, 2017.

\bibitem{intro2}
M.~C. Caccami, M.~Y.~S. Mulla, C.~Occhiuzzi, C.~Di~Natale, and G.~Marrocco, ``{Design and Experimentation of a Batteryless On-Skin RFID Graphene-Oxide Sensor for the Monitoring and Discrimination of Breath Anomalies},'' \emph{IEEE Sensors Journal}, vol.~18, no.~21, pp. 8893--8901, 2018.

\bibitem{intro3}
X.~Liu, J.~Yin, Y.~Liu, S.~Zhang, S.~Guo, and K.~Wang, ``{Vital Signs Monitoring with RFID: Opportunities and Challenges},'' \emph{IEEE Network}, vol.~33, no.~4, pp. 126--132, 2019.

\bibitem{intro4}
R.~Zhao, D.~Wang, Q.~Zhang, H.~Chen, and A.~Huang, ``{CRH: A Contactless Respiration and Heartbeat Monitoring System with COTS RFID Tags},'' in \emph{2018 15th Annual IEEE International Conference on Sensing, Communication, and Networking (SECON)}, 2018, pp. 1--9.

\bibitem{tiradar}
\BIBentryALTinterwordspacing
T.~Instruments, ``{MmWave radar sensors}.'' [Online]. Available: \url{https://www.ti.com/sensors/mmwave-radar/overview.html}
\BIBentrySTDinterwordspacing

\bibitem{intro5}
F.~Adib, H.~Mao, Z.~Kabelac, D.~Katabi, and R.~C. Miller, ``{Smart homes that monitor breathing and heart rate},'' in \emph{Proceedings of the 33rd Annual {ACM} Conference on Human Factors in Computing Systems}.\hskip 1em plus 0.5em minus 0.4em\relax New York, NY, USA: ACM, Apr. 2015.

\bibitem{Nature_paper}
M.~Mercuri, I.~Lorato, Y.~Liu, and et~al., ``{Vital-sign monitoring and spatial tracking of multiple people using a contactless radar-based sensor},'' \emph{Nat Electron}, no.~2, p. 252–262, 2019.

\bibitem{survey_VS}
J.~C. Soto, I.~Galdino, E.~Caballero, V.~Ferreira, D.~Muchaluat-Saade, and C.~Albuquerque, ``{A survey on vital signs monitoring based on Wi-Fi CSI data},'' \emph{Computer Communications}, vol. 195, pp. 99--110, 2022.

\bibitem{breathfinding}
N.~Patwari, L.~Brewer, Q.~Tate, O.~Kaltiokallio, and M.~Bocca, ``{Breathfinding: A Wireless Network That Monitors and Locates Breathing in a Home},'' \emph{IEEE Journal of Selected Topics in Signal Processing}, vol.~8, no.~1, pp. 30--42, 2014.

\bibitem{wifi1}
H.~Abdelnasser, K.~Harras, and M.~Youssef, ``{UbiBreathe: A Ubiquitous non-Invasive WiFi-based Breathing Estimator},'' 05 2015.

\bibitem{wifi2}
R.~Ravichandran, E.~Saba, K.-Y. Chen, M.~Goel, S.~Gupta, and S.~N. Patel, ``Wibreathe: Estimating respiration rate using wireless signals in natural settings in the home,'' in \emph{Pervasive Computing and Communications (PerCom), 2015 IEEE International Conference on}.\hskip 1em plus 0.5em minus 0.4em\relax IEEE, 2015, pp. 131--139.

\bibitem{phasebeat}
D.~Zhang, Y.~Hu, Y.~Chen, and B.~Zeng, ``{BreathTrack: Tracking Indoor Human Breath Status via Commodity WiFi},'' \emph{IEEE Internet of Things Journal}, vol.~6, no.~2, pp. 3899--3911, 2019.

\bibitem{breathtrack}
X.~Wang, C.~Yang, and S.~Mao, ``{PhaseBeat: Exploiting CSI Phase Data for Vital Sign Monitoring with Commodity WiFi Devices},'' in \emph{2017 IEEE 37th International Conference on Distributed Computing Systems (ICDCS)}, 2017, pp. 1230--1239.

\bibitem{holo}
A.~Eid, J.~Zhu, L.~Xu, J.~G.~D. Hester, and M.~M. Tentzeris, ``{Holography-Based Target Localization and Health Monitoring Technique Using UHF Tags Array},'' \emph{IEEE Internet of Things Journal}, vol.~8, no.~19, pp. 14\,719--14\,730, 2021.

\bibitem{mtrack}
D.~Zhang, Y.~Hu, and Y.~Chen, ``{MTrack: Tracking Multiperson Moving Trajectories and Vital Signs With Radio Signals},'' \emph{IEEE Internet of Things Journal}, vol.~8, no.~5, pp. 3904--3914, 2021.

\bibitem{chenglonglocalization}
\BIBentryALTinterwordspacing
C.~Li, S.~De~Bast, Y.~Miao, E.~Tanghe, S.~Pollin, and W.~Joseph, ``{Contact-Free Multi-Target Tracking Using Distributed Massive MIMO-OFDM Communication System: Prototype and Analysis},'' 2022. [Online]. Available: \url{https://arxiv.org/abs/2208.10863}
\BIBentrySTDinterwordspacing

\bibitem{radar8}
M.~Mercuri, I.~R. Lorato, Y.-H. Liu, F.~Wieringa, C.~Van~Hoof, and T.~Torfs, ``Vital-sign monitoring and spatial tracking of multiple people using a contactless radar-based sensor,'' \emph{Nature Electronics}, vol.~2, no.~6, pp. 252--262, Jun. 2019.

\bibitem{radar771}
A.~Ahmad, J.~C. Roh, D.~Wang, and A.~Dubey, ``{Vital signs monitoring of multiple people using a FMCW millimeter-wave sensor},'' in \emph{2018 IEEE Radar Conference (RadarConf18)}, 2018, pp. 1450--1455.

\bibitem{radar77}
Z.~Xu, C.~Shi, T.~Zhang, S.~Li, Y.~Yuan, C.-T.~M. Wu, Y.~Chen, and A.~Petropulu, ``{Simultaneous Monitoring of Multiple People’s Vital Sign Leveraging a Single Phased-MIMO Radar},'' \emph{IEEE Journal of Electromagnetics, RF and Microwaves in Medicine and Biology}, vol.~6, no.~3, pp. 311--320, 2022.

\bibitem{radar60}
T.~Sakamoto, ``{Noncontact Measurement of Human Vital Signs during Sleep Using Low-power Millimeter-wave Ultrawideband MIMO Array Radar},'' in \emph{2019 IEEE MTT-S International Microwave Biomedical Conference (IMBioC)}, vol.~1, 2019, pp. 1--4.

\bibitem{devalg}
\BIBentryALTinterwordspacing
L.~J. Dirksmeyer, A.~Marnach, D.~Schmiech, and A.~R. Diewald, ``{Developing of Algorithms Monitoring Heartbeat and Respiration Rate of a Seated Person with an FMCW Radar},'' \emph{Advances in Radio Science}, vol.~19, pp. 195--206, 2021. [Online]. Available: \url{https://ars.copernicus.org/articles/19/195/2021/}
\BIBentrySTDinterwordspacing

\bibitem{alidoustaghdam2023enhancing}
H.~Alidoustaghdam, M.~Chen, B.~Willetts, K.~Mao, A.~Kokkeler, and Y.~Miao, ``{Enhancing Vital Sign Estimation Performance of FMCW MIMO Radar by Prior Human Shape Recognition},'' 2023.

\bibitem{testbed1}
X.~Wang, M.~Laabs, D.~Plettemeier, K.~Kosaka, and Y.~Matsunaga, ``{28 GHz Multi-Beam Antenna Array based on Wideband High-dimension 16x16 Butler Matrix},'' in \emph{2019 13th European Conference on Antennas and Propagation (EuCAP)}, 2019, pp. 1--4.

\bibitem{testbed4}
R.~Hersyandika, Q.~Wang, S.~Pollin, Y.~Miao, and F.~Tufvesson, ``{Measurement-based Analysis of mmWave Multi-Point Connectivity Approaches in Blocked Scenarios},'' 2022.

\bibitem{testbed2}
A.~Colpaert, E.~Vinogradov, and S.~Pollin, ``{Fixed mmWave Multi-User MIMO: Performance Analysis and Proof-of-Concept Architecture},'' in \emph{2020 IEEE 91st Vehicular Technology Conference (VTC2020-Spring)}, 2020, pp. 1--5.

\bibitem{testbed3}
\BIBentryALTinterwordspacing
N.~Instruments, ``{5G massive MIMO testbed: From theory to reality},'' 2021. [Online]. Available: \url{https://www.ni.com/en-us/innovations/white-papers/14/5g-massive-mimo-testbed--from-theory-to-reality--.html}
\BIBentrySTDinterwordspacing

\bibitem{ni}
\BIBentryALTinterwordspacing
``Labview communications mimo application framework 19.5 readme.'' [Online]. Available: \url{https://www.ni.com/pdf/manuals/377191e.html}
\BIBentrySTDinterwordspacing

\bibitem{mobi}
\BIBentryALTinterwordspacing
TMSi, ``{User manual: Mobi}.'' [Online]. Available: \url{https://info.tmsi.com/user-manual-mobi}
\BIBentrySTDinterwordspacing

\bibitem{10081872}
M.~Thoonen, P.~Veltink, F.~Halfwerk, R.~Van~Delden, and Y.~Wang, ``{A Movement-Artefact-Free Heart-Rate Prediction System},'' in \emph{2022 Computing in Cardiology (CinC)}, vol. 498, 2022, pp. 1--4.

\bibitem{csi1}
X.~Wang, C.~Yang, and S.~Mao, ``{PhaseBeat: Exploiting CSI Phase Data for Vital Sign Monitoring with Commodity WiFi Devices},'' in \emph{2017 IEEE 37th International Conference on Distributed Computing Systems (ICDCS)}, 2017, pp. 1230--1239.

\bibitem{datacal1}
C.~Li, S.~De~Bast, E.~Tanghe, S.~Pollin, and W.~Joseph, ``{Toward Fine-Grained Indoor Localization Based on Massive MIMO-OFDM System: Experiment and Analysis},'' \emph{IEEE Sensors Journal}, vol.~22, no.~6, pp. 5318--5328, 2022.

\bibitem{datacal2}
M.~Speth, S.~Fechtel, G.~Fock, and H.~Meyr, ``Optimum receiver design for wireless broad-band systems using ofdm. i,'' \emph{IEEE Transactions on Communications}, vol.~47, no.~11, pp. 1668--1677, 1999.

\bibitem{datacal3}
A.~Tarighat, R.~Bagheri, and A.~Sayed, ``{Compensation schemes and performance analysis of IQ imbalances in OFDM receivers},'' \emph{IEEE Transactions on Signal Processing}, vol.~53, no.~8, pp. 3257--3268, 2005.

\bibitem{smooth}
\BIBentryALTinterwordspacing
Mathworks, ``smooth.'' [Online]. Available: \url{https://nl.mathworks.com/help/curvefit/smooth.html}
\BIBentrySTDinterwordspacing

\bibitem{gan}
M.~Gan, \emph{\BIBforeignlanguage{English}{{ Accurate and low-complexity ray tracing channel modeling}}}.\hskip 1em plus 0.5em minus 0.4em\relax TU Wien, 2015.

\bibitem{roughness}
S.~Ritter, K.~Staub, and P.~Eppenberger, ``{Associations between relative body fat and areal body surface roughness characteristics in 3D photonic body scans—a proof of feasibility},'' \emph{International Journal of Obesity}, vol.~45, pp. 1--8, 04 2021.

\bibitem{dwt}
S.~Sardy, P.~Tseng, and A.~Bruce, ``Robust wavelet denoising,'' \emph{IEEE Transactions on Signal Processing}, vol.~49, no.~6, pp. 1146--1152, 2001.

\bibitem{lung}
J.~F. Murray, \emph{\BIBforeignlanguage{English}{{The normal lung : the basis for diagnosis and treatment of pulmonary disease / John F. Murray}}}.\hskip 1em plus 0.5em minus 0.4em\relax Saunders Philadelphia, 1976.

\bibitem{rloss}
T.~Wu, T.~S. Rappaport, and C.~M. Collins, ``{The human body and millimeter-wave wireless communication systems: Interactions and implications},'' in \emph{2015 IEEE International Conference on Communications (ICC)}, 2015, pp. 2423--2429.

\end{thebibliography}

\end{document}